\documentclass[preprint,authoryear,12pt]{elsarticle}

\usepackage{graphics}

\usepackage{amssymb}

\journal{Journal of Atmospheric and Solar-Terrestrial Physics}

\begin{document}

\begin{frontmatter}



\title{Waves in vertically inhomogeneous dissipative atmosphere}

\author[label1]{I.S. Dmitrienko}
\ead{dmitrien@iszf.irk.ru}
\author[label2]{G.V. Rudenko}
 \ead{rud@iszf.irk.ru}

\address[label1]{Institute of Solar-Terrestrial Physics SB RAS, Irkutsk, Russia}
\address[label2]{Institute of Solar-Terrestrial Physics SB RAS, Irkutsk, Russia}

\begin{abstract}
A method of construction of solution for acoustic-gravity waves
(AGW) above a wave source, taking dissipation throughout the
atmosphere into account (Dissipative Solution above Source, DSAS),
is proposed. The method is to combine  three solutions for three
parts of the atmosphere:  an analytical solution for the upper
isothermal part and  numerical  solutions  for the real
non-isothermal dissipative atmosphere in the middle part and for
the real non-isothermal small dissipation atmosphere  in the lower
one. In this paper the method has been  carried out for the
atmosphere with  thermal conductivity but without viscosity. The
heights of strong dissipation and the total absorption index  in
the regions of weak and average dissipation are found. For
internal gravity waves the results of test calculations for an
isothermal atmosphere and calculations for a real non-isothermal
atmosphere are shown in graphical form. An algorithm and
appropriate code to calculate DSAS, taking dissipation due to
finite thermal conductivity into account  throughout the
atmosphere,  are developed. The results of test DSAS calculations
for an everywhere isothermal atmosphere are given. The calculation
results for DSAS for the real non-isothermal atmosphere are also
presented. A method for constructing of the 2x2 Green's matrix
fully taking dissipation into account and  allowing  to find
disturbance from  some source  of  AGW  in  the atmosphere is
proposed.
\end{abstract}


\end{frontmatter}


\section{Introduction}\label{section1}
The study of acoustic-gravity waves (AGW)  in the atmosphere has a
long history. The basic properties of AGW  propagation were
summarized even in  \cite{Hook}. However, dissipationless approach
limits the applicability of the results. They apply only to the
waves in the lower and middle atmosphere, where weakness of
dissipation allows to neglect it. High enough in the upper
rarefied atmosphere  dissipation becomes the determining factor in
the behavior AGW. Therefore, to describe the penetration of AGW in
the upper atmosphere it is necessary to involve more complex
formalism of the dissipative hydrodynamics. In the dissipative
hydrodynamics  an analytical description of AGW is obtained in the
frames of the isothermal approximation  \citep{Lyons, Yanowitch_a,
Yanowitch_b, Rudenko_a,Rudenko_b}. However, isothermal
approximation models  the real atmosphere properties adequately in
its upper layers only. Without the isothermal approximation the
wave dissipation was taken into account in \cite{Vadas}. These
researches were performed with use of the WKB approximation, which
has its limitations. It enables to take into account the weak
dissipation in the lower and middle layers of the atmosphere and
does not allow to adequately describe the strong effect of
dissipation on the wave in the upper atmospheric layers.

This paper is devoted to an AGW solution fully taking dissipation
into account. The solution is constructed as an integrated whole
for two fundamentally different physical conditions of the
atmosphere, with virtually no dissipation in the lower and middle
atmosphere and with substantial dissipation, increasing with the
height, in the upper atmosphere. The solution has two parameters:
real wave frequency and horizontal wave number.

The absence of the flow of energy towards the Earth in the upper
layers of the atmosphere is the only boundary condition for the
solution; because of dissipativity  of the atmosphere, it
coincides with the condition of upward attenuation of the solution
(from the Earth). The solution exists at all  heights. Formally,
it can be taken as a solution of the wave problem with some source
on the Earth surface. However, wider physical meaning  of this
solution is that the part of this solution, with  real wave
frequency and horizontal wave number, in the region above any
source located in the atmosphere (or on the Earth's surface),
describes the vertical structure of the disturbance produced  by
this source there in this region . Region above a source is a part
of the atmosphere, the heights of which exceed the upper limit of
the source localization; there are no limits for horizontal
coordinates of this region. Formally, the solution extends
indefinitely in the upper atmosphere. Starting from some height it
decreases against the rarefying atmosphere background due to
increase of the effect of dissipation with altitude. Given
dissipative character  of the solution and its aforementioned
physical meaning we call it dissipative solution above source
(DSAS).

DSAS is necessary final  element for the solution of  the problem
of the wave propagation  from  a source located in the atmosphere.

Now  there are well-developed methods for description of wave
phenomena in the real atmosphere, based on the direct numerical
solution of hydrodynamic equations, which include a variety of
model sources of climate and  anthropogenic character:
\citep{Hickey1997, Hickey1998, Walterscheid1990, Walterscheid2001,
Snively2003, Snively2005, Snively2007, Yu2007_a,
Yu2007_b,Yu2007_c, Yu2009, Kshevetskii2005, Gavrilov2014}. These
methods are based on direct numerical integration of the nonlinear
system of partial differential equations in two- or
three-dimensional approximations for stratified atmosphere with
viscosity and thermal conductivity and wind stratification.
However, these methods are inefficient when describing the
perturbation at large distance from the localization of the
source, because of their resource consumption and rapid
accumulation of numerical errors in the need to increase the grid
to ensure sufficient spatial resolution. It is well known that the
wave disturbances, propagating far from their sources, are
important elements of the dynamics of the atmosphere. Due to
increase of their relative amplitudes in height, they manifest
themselves in the thermosphere as the main cause of the diverse
ionospheric disturbances. The transfer of wave energy and momentum
over large distances is provided due to presence of the conditions
of waveguide propagation in the temperature inhomogeneities in the
lower and middle atmosphere \cite{Pierce1970}. The most effective
method  to describe the remote signals is to present them as
superpositions of the waveguide modes generated by the source (the
method of the normal modes). An algorithm, for solving the
disturbance propagation problem, based on the method of normal
modes, described in detail in \cite{Pierce1971}, where a result of
comparison of the calculated and the observed pressure
disturbances on the Earth's surface in the far zone of a nuclear
explosion was first presented. In principle, based on the
algorithm of \cite{Pierce1971}, synthesizing propagating normal
modes, one can calculate the disturbance parameters at any height
where the dissipation-free approach, in which \cite{Pierce1971},
built its own algorithm,  still holds.

Our main goal is to study the possibility of similar description
of the  long-range propagation of disturbances in the ionospheric
heights. In the ionosphere, according to the basic concept
\cite{Hines1960}, the so-called travelling ionospheric
disturbances (TIDs) are the result of infiltration of  the neutral
atmosphere disturbances from the lower IGW (internal gravity wave)
waveguide channels. In the upper atmosphere in the tropospheric
heights  the dissipative effect on the wave processes becomes
dominant that makes it impossible to ignore and even to take into
account approximately the dissipative terms of the wave equations
system. As a result, we have to deal with  a significantly more
complicate problem compared to that of \cite{Pierce1971}.

As in the dissipation-free case, \cite{Pierce1971}, the key
element of the disturbance propagation problem is to find the
Green's function (or matrix), for the construction of which we
need two solutions: the lower one (below the source) and the upper
one (above the source) satisfying the lower and upper boundary
conditions, respectively. In the case, when a source is in the
range of altitudes where the dissipation-free approximation is
applicable, then finding a solution under source can now be
considered as a routine problem. Thus, in this case, the key
problem is to find a solution above the source for a realistic
model of the atmosphere with a full account of dissipation. It is
the problem the present paper focuses on. Our approach to this
problem bases on matching a solution of the weakly-dissipative
wave problem of second order for the lower part of the atmosphere
with a solution of the dissipative wave problem of order, higher
than second, for the upper part of the atmosphere. It is different
essentially, on the level of severity, from the method proposed in
its time in \cite{Francis1973_a}. In the paper
\cite{Francis1973_a}, a solution over the source was obtained with
the so-called multi-layer technique, in which the medium in each
horizontal layer is considered isothermal with constant
coefficients of viscosity and thermal conductivity; so
representation of the solution in each such layer is possible in
analytical form. When taking dissipation  into account,  there
are, generally speaking, three wave types with three complex
values of the square of the vertical wave number, corresponding to
(in the weak dissipation case) conventional acoustic-gravity wave
(AGW)  and two types of purely dissipative oscillations caused by
viscosity and thermal conductivity. \cite{Francis1973_a}
represented the solution in each layer only with AGV solutions,
and only two conditions of six ones used for matching solutions of
adjacent layers. Such an approach is quite justified in the
temperate heights, where dissipation is weak, but obviously
incorrect in the strong  dissipation regions. Thus, in order to
solve the problem, Francis artificially lowered the order of the
system of equations from the 6th  order up to the second order
everywhere.  Thus, the structure of wave disturbances in the lower
part of the atmosphere and the dispersion characteristics of the
modes, captured by the inhomogeneities of the lower atmosphere,
the method of Francis allows to calculate sufficiently well, but
it obviously is not able to give a correct description of
disturbances in the upper atmosphere. The  results of Francis are
widely used  in the theoretical researches and in the
interpretation of observations of various disturbances, including
those in the upper atmosphere \citep{Shibata1983, Afraimovich2001,
Vadas2009, Vadas2012, Idrus2013, Heale2014, Hedlin2014}. Unlike
Francis, we use a method of reducing the order of the system  of
the wave equations up to the second one (in our own version) only
for the small-dissipation altitudes where it is justified.
Therefore, our method, unlike the method of Francis, gives the
possibility to describe disturbances of the upper atmosphere
adequately. Another approach to obtain the solution above source
is developed in the papers \citep{Lindzen1970, Lindzen1971_a,
Lindzen1971_b}. This approach, based on the numerical method of
the  tridiagonal matrix algorithm for the full system of linear
dissipative hydrodynamic equations, can be considered quite
strict.  The only restriction is the circumstance that the
hydrostatic approximation is used, which is valid only for
sufficiently long-period oscillations  (with frequencies much
lower  than  the Brunt-V\"{a}is\"{a}l\"{a} frequency). The results
obtained on the basis of this approach  played an important role
in the description of tidal modes and continue to be relevant
until now \citep{Forbes1979, Fesen1995, Gavrilov1995,
Grigorev1999, Akmaev2001, Angelatsi2002, Yu2009}.

Thus, our approach to obtaining the solution above source  differs
from  \cite{Francis1973_a} by correct consideration  of
dissipation; as for the papers on the base of \cite{Lindzen1970},
it is of more universal character since does not have restrictions
caused  by usage of the hydrostatic approximation. It should be
noted that the dissipation is taken into account in our present
paper less complete than in the aforementioned papers. This is due
to the fact that we believe it possible, as a first step, consider
the viscous dissipation as negligible, taking  only the
dissipation due to the heat conductivity into account. We use such
an approach in connection with the possibility of a simple
analytic representation of solutions in heat-conducting
thermosphere, which allows us to put the upper boundary conditions
more strict at the infinity. Such conditions are more accurate
compared with the boundary conditions at some height used in
\cite{Lindzen1970}. It is also a useful approach as giving  a
clear understanding of dissipation influence on  the wave motion
in the upper atmosphere.

Because in reality the dissipation, associated with the thermal
conductivity, begins  its influence  at  lower altitude then the
viscous dissipation does, our model of the atmosphere can give the
real picture of the disturbances in the upper atmosphere
sufficiently well, at least, to a certain height. As will be
discussed below, taking into account the atmosphere viscosity is,
in principle, possible in our approach to finding DSAS. It is also
feasibly  to  include, in the model,  the atmosphere wind, which
is not considered in this paper.

DSAS formally exists for the entire atmosphere up to the Earth and
corresponds in this form (for all heights) to the particular case
of a source on the Earth's surface.If such a solution for certain
values of the frequency   and horizontal wave number $k_x$
satisfies the lower boundary condition, it can be regarded as a
solution for a trapped mode.

In general case with  the wind, the continuation of DSAS up to the
Earth is possible  in such cases when the horizontal phase
velocity of the wave is large  enough so it is not equal to the
wind speed at any height. Otherwise, because of the so-called wave
destruction (the phenomenon, responsible for the turbulent
properties of the real atmosphere), \cite{Lindzen1981}, DSAS will
be applied only to a certain height and can be used for cases of
sources at sufficiently high altitudes. However, the horizontal
phase velocities of the waveguide modes,  which are responsible
for the distant propagation \citep{Francis1973_b}, are of the
order of the speed of sound, so the condition of equality of the
phase velocity to the wind speed never is satisfied.

It should also be noted that the feature of our method is absence
of the need to include in the equations the additional dissipative
terms taking into consideration the impact of turbulence in a
model manner, without which none of the previously developed
methods can work, because of the inherent numerical instability.

DSAS with a full view of the dissipation is most important for the
synthesis of the waveforms at fairly high altitudes, where it is
impossible to consider dissipation with any approximate method.
Such altitudes are typical for TIDs. To synthesize waveforms at
lower altitudes the algorithms for non-dissipative atmosphere are
sufficient, for example, the above algorithm of \cite{Pierce1971}.
Note that the wave forms of the disturbances in the upper
atmosphere should be less complicated than those in the middle and
lower altitudes, due to the dissipative suppression of multiple
waveguide modes of high-frequency sound range. As shown by
\cite{Francis1973_b}, only one or two major IGW modes can take
place at ionospheric heights, due to the waveguide leakage and the
weak horizontal attenuation.

Our consideration, as well as the general problem of the synthesis
of waveforms by normal modes, requires linearity of disturbance at
all altitudes. Taking dissipation into account provides drop with
height of the disturbance amplitude, relative to the background
parameters, in the upper atmosphere from a certain level. This
circumstance enables the selection of source amplitude small
enough for the disturbance to be linear throughout. At first
glance, it may seem that limitations of the source amplitude, to
ensure linearity of the disturbance, have to be very strong and so
the real sources unlikely satisfy them - since the dissipative
suppression of growth of the relative amplitude of the disturbance
with height, as a rule, begins at the altitudes, at which the
disturbance amplitude can grow by orders of magnitude  compared to
the amplitude near the Earth. Indeed, the disturbances have likely
to be non-linear just above the typical sources localized in space
and time. However, we must take into account the fact that only a
small part of the disturbance energy is captured by the
low-atmosphere waveguides, capable of transferring the wave energy
over long distances. In addition,  if the main source spectral
power is distributed in the high-frequency range (this is typical,
for example, for explosions), then the amplitudes  of the
low-frequency spectrum range,  generating waveguide modes, will be
significantly less than the main amplitudes of the source. All
these factors quite can lead to linearity of the waveguide modes.

Note,  in particular,  the fact that TIDs are typically  very
close to harmonical waves \citep{Hook} says also in favor of the
assumption of linearity of the actual responses in the upper
atmosphere. Thus, we can assume that the linear solution above the
source can be the basis for solving the propagation problem for
disturbances in a stratified atmosphere,  if the amplitude of the
source is small enough, or the ducting propagation only is of
interest.

The solution to this problem for a stratified atmosphere within
the dissipationless approximation is well known \cite{Hook}. In
this case, we have the second-order problem, i.e. it is formulated
as a second-order differential equation or a set of two
first-order differential equations equivalent to it. To solve the
dissipationless  problem with a source the Green's matrix composed
of two second-order problem  solutions is used. One of these
solutions must satisfy the upper boundary condition and the other
must satisfy the condition on the Earth. It is shown in this paper
that one can also limit to a second-order problem in the small
dissipation approximation for disturbance of not too small
vertical scale. Therefore, the solution of the problem with a
source is also obtained in this case by means of the second-order
Green's matrix. But the small dissipation approximation is
applicable only in the lower and middle parts of the atmosphere.
It is not sufficient in the upper part because of increase with
altitude  of the dissipation effect. So we have to solve a problem
of order higher, then second,  there.

But even fully taking dissipation into account, one can retain a
formalism of a second-order problem to a significant extent if the
source is located in the altitude range where dissipation is
small.  For this it is enough to replace the solution of the
second-order problem in the $2\times 2$ Green's matrix that
satisfies the upper boundary condition with DSAS. The Green's
matrix obtained in such a way we will call the extended Green's
matrix. Note that in the atmosphere mainly occur sources for which
the method of the extended Green matrix is applicable. These are
sources located in the lower and middle atmosphere: EXPLOSIONS
tsunami meteorites anthropogenic sources, meteorological
phenomena.

Description of an extended tail formally indefinitely outgoing to
the upper atmosphere is the main feature of DSAS. However, if a
wave is attenuated too much for its extension to the upper
atmosphere, the amplitude of the upper atmospheric tail can be so
small that the disturbance with such an amplitude is of no
interest. On the other hand, if a wave weakly damps when
propagating to the upper atmosphere, the relative amplitude of the
tail (the ratio of the amplitude to the background) can be too
large in the upper atmosphere for the disturbance to be under the
linear approximation. In this connection, we investigated effect
of dissipation on the AGV propagation. Some characteristics of
the wave dissipative loss are introduced and, then, their values
are calculated.

We organize our paper as follows. Equations used further to
construct a DSAS are derived in Section 2. We have demonstrated
the possibility of obtaining from the hydrodynamic equations both
a system of four first-order ODEs taking into account  thermal
conductivity only and, in more general case, a system of six
first-order ODEs assuming, besides consideration of thermal
conductivity, atmospheric viscosity.  Numerical solutions of such
systems can be obtained by the standard Runge-Kutta method, if the
initial values of  the unknowns are given at the specified height.
The possibility of including wind stratification in the model of
the atmosphere is indicated.  To describe waves in the weakly
dissipative (lower) part of the atmosphere we obtained a system of
two first-order ODE.

Section 3 gives details on the method for an analytical solution
for the upper atmosphere with an isothermal temperature
distribution. The theory is based on the possibility of describing
wave disturbances in an isothermal atmosphere by one sixth-order
differential equation in the general case of simultaneous
consideration of viscosity and thermal conductivity or
fourth-order equation if we consider the thermal-conductivity
model of the atmosphere without viscosity. In this paper, we have
constructed DSAS for the case of the thermal-conductivity model of
the atmosphere without viscosity. In this case, a fourth-order
equation reduces to the generalized hypergeometric differential
equation \citep{Lyons, Yanowitch_a, Yanowitch_b,
Rudenko_a,Rudenko_b} analytical solutions of which are
subsequently used for the isothermal part of the atmosphere. In
section 3, we also study the effect of dissipation on the AGV
propagation. First,  the parameter $z_c$, characterizing the
heights of the transition to the regime of strong dissipation, is
calculated. This parameter is singled out when deriving  the
hypergeometric differential equation  \citep{Lyons, Yanowitch_a,
Yanowitch_b, Rudenko_a,Rudenko_b}. In the heights below $z_c$, we
have approximately classical oscillations corresponding to usual
wave types of oscillations without dissipation. In the range of
heights higher $z_c$, dissipation completely changes nature of
oscillations transforming them into purely dissipative waves. It
follows from the equations that $z_c$ depends only on the wave
period for any vertical profile of the atmospheric density, so we
calculated it as a function  of period. The parameter $z_c$,
calculated for the isothermal atmosphere, can be used for
estimating of the height of the strong dissipation region,
depending on wave period, in the real atmosphere. Second, as a
dissipation characteristic at the heights below $z_c$, we
introduced an index of total vertical dissipation . We calculated
it both on the basis of the WKB-approximation solution for the
non-isothermal  thermal-conductive atmosphere and and the
analytical solution for isothermal  thermal-conductive atmosphere.

Section 4 gives our method for constructing of DSAS. We  have
carried  out this method in the frames of the thermal-conductivity
model of the atmosphere without viscosity. In this case, the
complete set of wave equations is presented by four first-order
ordinary differential equations which, as we show, can be
numerically solved in the upper part of the atmosphere, where
dissipation is strong, to some height in the middle atmosphere,
where dissipation effect is sufficiently small. There are
calculation difficulties not allowing us to use a complete set of
wave equations in the lower part of the atmosphere. On the other
hand, dissipation is small there. This allowed us to reduce the
fourth-order problem in the lower part of the atmosphere to a
second-order problem. Thus,  DSAS is constructed of the solutions
for three height ranges in the direction from the upper part of
the atmosphere to the Earth: an analytical solution for the
isothermal upper atmosphere (I); a numerical solution of the
system of four ODEs in a non-isothermal upper atmosphere until the
height of $z_t$ on which the dimensionless kinematic thermal
conductivity reaches a sufficient smallness (II); a solution of
the system of two ODEs for the small dissipation approximation in
the region below  $z_t$ up to the Earth's surface (III). We will
further also call the height ranges I, II, III lower middle and
upper  parts of the atmosphere, respectively.

An analytical solution (I), satisfying the condition of upward
attenuation, is a superposition of a superposition of some two
solutions, having asymptotics each of which attenuates upward; the
coefficients of these solutions are arbitrary.  DSAS is continuous
at the boundary of the regions (I) and (II) because an analytical
solution (I) at this boundary gives four  initial values for
solving the system of four ODEs numerically in the region (II).

As for the boundary of the regions (II) and (III) then  we applied
our method for matching the solutions (II) and (III) there.
Matching solutions (II) and (III) is important element in
constructing  DSAS. It is clear that the solutions corresponding
to different approximations can not be match arbitrarily  closely
at the boundary. However, it is also clear that this is not
required, the smallness of jumps is enough. Therefore matching the
solutions (II) and (III)  carries out    as follows. We take
values of any two components from four components of the solution
(II) at the boundary of the regions (II) and (III) (the height
$z_t$) as initial values for solving the system of two ODEs
numerically in the region (II). For two other components, we
provide sufficient smallness of the jump at the height $z_t$,using
the condition of minimization  of the jumps to find  the ratio
between the unknown coefficients in the solution (I) . For the
minimal jumps would be sufficiently small, a sufficiently small
height $z_t$ must be chosen. Influence of $z_t$ values on the
values of the jumps is analyzed using special calculations.

We test our method for constructing of DSAS, using an everywhere
isothermal model of the atmosphere, in section 5, and we give the
result of calculation of DSAS for a real non-isothermal model of
the atmosphere in section 6. Finally, in section 7, we give the
extended Green's matrix, allowing to describe the propagation of
disturbances in the atmosphere from the various sources.

\section{Basic wave equations in a vertically stratified horizontally homogeneous non-isothermal atmosphere}\label{section2}
In the approximation of the stationary horizontally homogeneous
atmosphere, linear disturbances can be represented by a
superposition of waves of the form $f\left(z,\omega,\bf
{k}_{\perp} \right)e^{i\omega t+ i\left(\bf {k}_{\perp} \cdot \bf
{r} \right)}$, where: $z$ is the vertical coordinate,
$\left(\omega,\bf {k}_{\perp} \right)$ are the frequency and
horizontal wave number. As we will show, vertical structure of
such waves is described by either a system of $n$ first-order
ordinary differential equations with $n$ independent values or one
$n$-order ordinary differential equation for one value; $n$ is
two, four or six, depending on choice of approximation. Without
dissipation, we have the wave problem of the second order $(n=2)$
describing classical wave oscillations, for example, acoustic or
IGWs. With  taking into consideration  dissipation due to thermal
conductivity, the order of the wave problem becomes higher
$(n=4)$; two more wave solutions arise as a consequence. Adding
viscosity leads to $n=6$ and two more "dissipation" solutions. At
the problem order $n=2$, it is possible to reduce it to one
second-order differential equation \citep{Ost, Pon}. At all the
orders of the problem, it shall be  shown that it reduces to an
system of $n$ explicit first-order ordinary differential equations
(normal system).

We will give derivations of necessary equations for the problems
with different $n$. We will introduce all physical parameters in
the form of sums of their undisturbed and disturbed values: $f \to
f_0+f$. We will regard the system of hydrodynamic equations as
input:
 \begin{equation}\label{eq1}
\begin{array}{l}
  \partial_t\rho+\left({\bf v}_{0\perp}\cdot\nabla \right)\rho+\rho_0\nabla\cdot{\bf v}+{\rho^\prime}_0v_z=0,  \\
  \partial_t T+\left({\bf v}_{0\perp}\cdot\nabla \right)T+(\gamma-1)T_0\nabla\cdot{\bf v}+{T^\prime}_0v_z=\frac{\kappa}{c_v\rho_0}\Delta T, \\
  \rho_0\partial_t{\bf v}+\rho_0\left({\bf v}_{0\perp}\cdot\nabla \right){\bf v}+\nabla p-\rho{\bf g}=\nu_1\Delta{\bf v}
  +\left(\frac{1}{3}\nu_3+\nu_2  \right)\nabla(\nabla\cdot{\bf v}),\\
  p=RT_0\rho+R\rho_0T.
  \end{array}
\end{equation}
  System of equations (\ref{eq1}) is sequentially mass-continuity equations, heat-balance equation, momentum equations, and equations of
  state; $ \rho, p, T,{\bf v}$ is the disturbed values of density, pressure, temperature, and velocity, accordingly; disturbed
  values corresponding to them with the index $_0$ are the functions of only the vertical coordinate $z$, the prime symbol denotes differentiation by
  $z$; ${\bf g}$ is the free fall acceleration vector $(0,0,-g)$; $R$ is the universal gas
  constant; $c_v$ is the specific heat at constant volume;
  $\kappa$ is the dynamic coefficient of thermal
  conductivity, $\nu_1$ and $\nu_2$ is the dynamic coefficients of the first and second viscosity; the
  subscript $_\perp$ denotes a horizontal part of the vector. Without loss of generality, we will put the coordinate
  axis $x$ directed along a horizontal wave vector ${\bf
  k}_\perp$: $(k_x,0,0)$. In this case, all the disturbed values do not depend on $y$. In addition, it is easy to show that $y$ disturbed
  velocity component is determined
  by its equation independent of other disturbed parameters of the
  environment. Therefore, we can ignore it, defining the disturbed velocity vector by only
  two ${\bf v}=(v_x,0,v_z)$. Replacing partial derivatives with respect
  to values:
  $\partial_t, \partial_x$ with the factors $-i\omega, ik_x$, respectively, and introducing a special notation for a frequency
  function in the moving reference system
  $\Omega(z)=\omega-k_xv_0x(z)$, we modify the input system of equations to:
 \begin{equation}\label{eq2}
\begin{array}{l}
  -i\Omega\rho+\rho_0\left(ik_xv_x+{v_z}^\prime\right)+{\rho^\prime}_0v_z=0,  \\
  -i\Omega T+(\gamma-1)T_0\left(ik_xv_x+{v_z}^\prime\right)+{T^\prime}_0v_z=\frac{\kappa}{c_v\rho_0}\left(-k^2_xT+T^{\prime\prime}\right), \\
  -i\Omega v_x+ik_x\rho^{-1}_0 p=\nu_1\left(-k^2_xv_x+v_x^{\prime\prime}\right)
  +\left(\frac{1}{3}\nu_3+\nu_2  \right)\left(-k^2_xv_x+ik_xv_z^\prime\right),\\
  -i\Omega v_z+\rho^{-1}_0 p^\prime+g\rho_0^{-1}\rho=\nu_1\left(-k^2_xv_z^\prime+v_z^{\prime\prime}\right)
  +\left(\frac{1}{3}\nu_3+\nu_2  \right)\left(ik_xv_x+v_z^{\prime\prime}\right),\\
  p=RT_0\rho+R\rho_0T.
  \end{array}
\end{equation}
We will get the normal systems from (\ref{eq2}) in the form:
      \begin{equation}\label{eq3}
F'=\hat{A}F,
\end{equation}
where: $F$ is a vector of $n$ disturbance parameters sufficient to
formulate the wave problem in accordance with a chosen physical
approximation; $\hat{A}$ is the $n\times n$ matrix whose elements
depend on undisturbed environment parameters and the wave
parameters $(\omega,k_x)$

\subsection{Viscous heat-conducting atmosphere with wind $(n=6)$ (Case I)}\label{section2.1}
For this case, we will choose the following vector $F$  for the
system of equations (\ref{eq3}):
      \begin{equation}\label{eq4}
F=(T',T,v'_x,v_x,v'_z,v_z).
\end{equation}
To obtain a system of equations of the form (\ref{eq3}), we need
to have expressions for the second derivatives with respect to z
of the values $(T,v_x,v_z)$ through the components of vector $F$.
We get the equation for  $T''$ directly from the second equation
of system (\ref{eq2}). The equation for  $v'_x$ is obtained from
the third equation of system (\ref{eq2}), using the fifth and
first equations for expression of $p$. Furthermore,
differentiating the fifth and first equations of system
(\ref{eq2}), we find an expression for the combination $p'+\rho g$
containing only $v''_x$  and the vector  $F$ components.
Substituting this combination in the fourth equation, we get an
expression for  $v''_z$. As a result, we get for  $\hat A$:
 \begin{equation}\label{eq5}
\begin{array}{l}
  a_{11}=0; a_{12}=-i\Omega\frac{c_v \rho_0}{\kappa}+k^2_x; a_{13}=0; a_{14}=\frac{c_v \rho_0}{\kappa}(\gamma-1)T_0ik_x;  \\
  a_{15}=\frac{c_v \rho_0}{\kappa}(\gamma-1)T_0; a_{16}=\frac{c_v \rho_0}{\kappa}T'_0;\\
  a_{21}=1; a_{22}=0; a_{23}=0; a_{24}=0; a_{25}=0; a_{26}=0;\\
   a_{31}=0;  a_{32}=\frac{ik_xp_0}{\nu_1T_0}; a_{33}=0; a_{34}=\frac{ik_x^2p_0}{\Omega\nu_1}
   -\frac{i\Omega\rho_0}{\nu_1}+(\frac{4}{3}+\frac{\nu_2}{\nu_1})k_x^2;\\
   a_{35}=\frac{k_xp_0}{\Omega\nu_1}-(\frac{1}{3}+\frac{\nu_2}{\nu_1})ik_x;a_{36}=\frac{k_xp_0\rho'_0}{\Omega\nu_1\rho_0};\\
a_{41}=0; a_{42}=0; a_{43}=1; a_{44}=0; a_{45}=0; a_{46}=0;\\
 a_{51}=\frac{p_0}{\chi T_0};a_{52}=\frac{R\rho'_0}{\chi};
 a_{53}=-\frac{ik_x}{\chi}(\frac{ip_0}{\Omega}+\frac{1}{3}\nu_1+\nu_2);
 \\
 a_{54}=\frac{k_x}{\chi}\left(\frac{\rho_0}{\Omega}(g+RT'_0) +RT_0\left(\frac{\rho_0}{\Omega}\right)^\prime \right)
 ;\\
 a_{55}=\frac{1}{i\chi}\left(\frac{\rho_0}{\Omega}(g+RT'_0)+RT_0\left(\left(\frac{\rho_0}{\Omega}\right)^\prime +\frac{\rho'_0}{\Omega} \right)\right);\\
a_{56}=\frac{1}{i\chi}\left(\frac{\rho'_0}{\Omega}(g+RT'_0)+\Omega\rho_0+i\nu_1k_x^2+RT_0\left(\frac{\rho'_0}{\Omega}\right)^\prime\right);\\
a_{61}=0; a_{62}=0; a_{63}=0; a_{64}=0; a_{65}=1; a_{66}=0.
  \end{array}
\end{equation}
Here $\chi=\frac{4}{3}(nu_1+nu_2)+\frac{ip_0}{\Omega}$,
$p_0=R\rho_0T_0$. The variables $p$   and $\rho$  not belonging to
the vector $F$  are expressed through its components of the first
and fifth equations of system (\ref{eq2}):
 \begin{equation}\label{eq6}
\begin{array}{l}
  \rho=\frac{\rho_0}{i\Omega}(ik_xf_4+f_5)+\frac{\rho'_0}{i\Omega}f_6,\\
  p=RT_0\frac{\rho_0}{i\Omega}(ik_xf_4+f_5)+RT_0\frac{\rho'_0}{i\Omega}f_6+R\rho_0f_2.
  \end{array}
\end{equation}
Eqs. (\ref{eq3})-(\ref{eq6}) describe wave disturbances for the
most general case of the model of the atmosphere. With given
initial $F$ values, these equations can be used for numerical
solving to find corresponding wave disturbances. As we shall see
further, numerical obtaining of the solution to
(\ref{eq3})-(\ref{eq6}) is possible only when heights are
sufficiently great and correspond to not too high values of
undisturbed density.  Underneath, the problem is ill-conditioned.
We will show later that one can avoid this difficulty, using
smallness of dissipation at lower heights to reduce the problem to
the second one.
\subsection{Non-viscous heat-conducting atmosphere approximation $(n=4)$ (Case II)}\label{section2.2}
This approximation is main in this paper. As we will not take the
wind into account in this paper later, we will draw all the
calculations here without it. The formulas taking wind into
account, in this and following sections, are obtained by the
substitution  $\omega\to \Omega$. In the case of non-viscous
heat-conducting atmosphere, system of equations (\ref{eq2})
becomes:
\begin{equation}\label{eq7}
\begin{array}{l}
  -i\Omega\rho+\rho_0\left(ik_xv_x+{v_z}^\prime\right)+{\rho^\prime}_0v_z=0,  \\
  -i\Omega T+(\gamma-1)T_0\left(ik_xv_x+{v_z}^\prime\right)+{T^\prime}_0v_z=\frac{\kappa}{c_v\rho_0}\left(-k^2_xT+T^{\prime\prime}\right), \\
  -i\Omega v_x+ik_x\rho^{-1}_0 p=0,\\
  -i\Omega v_z+\rho^{-1}_0 p^\prime+g\rho_0^{-1}\rho=0,\\
  p=RT_0\rho+R\rho_0T.
  \end{array}
\end{equation}
To derive (\ref{eq7}) in the form of (\ref{eq3}), we choose the
following vector $F$ :
      \begin{equation}\label{eq8}
F=(T',T,p,v_z).
\end{equation}
We need expressions of the values $T'',p',v'_z$   through the
components of $F$. It is easy to get them excluding extra
variables $v_x$  and $\rho$  with use of the third and fifth
equation, accordingly. The expression for $T''$  is derived from
the second equation, using the expression $ik_xv_x+v'_z$  from the
first equation; the expression for  $p'$, from the fourth
equation; the expression for  $v'_z$, from the first equation. As
a result, we obtain the elements of the matrix  $\hat A$ in the
following form:
\begin{equation}\label{eq9}
\begin{array}{l}
  a_{11}=0;a_{12}=k_x^2-\frac{i\gamma\omega c_v\rho_0}{\kappa};a_{13}=\frac{i(\gamma-1)\omega c_v}{\kappa
  R};a_{14}=\frac{c_v\rho_0
  T_0}{\kappa}\left(\frac{T'_0}{T_0}-(\gamma-1)\frac{\rho'_0}{\rho_0}\right);\\
a_{21}=1;a_{22}=0;a_{23}=0;a_{24}=0;\\
a_{31}=0;a_{32}=\frac{g\rho_0}{T_0};a_{33}=-\frac{g}{RT_0};a_{34}=i\omega\rho_0;\\
a_{41}=0;a_{42}=-\frac{i\omega}{T_0};a_{43}=-\frac{i\omega}{\rho_0}\left(\frac{p'_0}{p_0g}+\frac{k_x^2}{\omega^2}\right);a_{44}=-\frac{\rho'_0}{\rho_0}.
  \end{array}
\end{equation}
The variables $v_x$  and  $\rho$ not belonging to the vector $F$
are expressed through its components from the third and fifth
equations of (\ref{eq7}):
 \begin{equation}\label{eq10}
\begin{array}{l}
  v_x=\frac{k_x}{\Omega}f_3,\\
  p=\frac{1}{RT_0}f_3-\frac{\rho_0}{T_0}f_2.
  \end{array}
\end{equation}
Eqs. (\ref{eq3}), (\ref{eq8})-(\ref{eq10}) describe wave
disturbances for the considered approximation. Also, as in the
previous case, these equations are applicable for numerical
calculations only for sufficiently large heights
\subsection{Dissipationless atmosphere approximation, $n=2$ (Case III)}\label{section2.3}
In this case, system of equations (\ref{eq2}) takes the following
form:
\begin{equation}\label{eq11}
\begin{array}{l}
  -i\Omega\rho+\rho_0\left(ik_xv_x+{v_z}^\prime\right)+{\rho^\prime}_0v_z=0,  \\
  -i\Omega T+(\gamma-1)T_0\left(ik_xv_x+{v_z}^\prime\right)+{T^\prime}_0v_z=0, \\
  -i\Omega v_x+ik_x\rho^{-1}_0 p=0,\\
  -i\Omega v_z+\rho^{-1}_0 p^\prime+g\rho_0^{-1}\rho=0,\\
  p=RT_0\rho+R\rho_0T.
  \end{array}
\end{equation}
We choose the following vector   to obtain an equation in the form
of (\ref{eq3}):
      \begin{equation}\label{eq12}
F=(p,v_z).
\end{equation}
\begin{equation}\label{eq13}
\begin{array}{l}
  a_{11}=\frac{p'_0}{\gamma p_0};
  a_{12}=i\omega\rho_0\left(1-\frac{\omega_N^2}{\omega^2}\right);\\
a_{21}=-\frac{i\omega}{\rho_0}\left(\frac{p'_0}{\gamma
p_0g}+\frac{k_x^2}{\omega^2}\right);a_{22}=-\frac{p'_0}{\gamma
p_0}.
  \end{array}
\end{equation}
Here  $\omega_N^2=-\frac{g^2}{c_s^2}-\frac{g\rho'_0}{\rho_0}$,
$c_s=\sqrt{\gamma RT_0}$ is sound velocity. $T,v_x,\rho$ and $T'$
are expressed through the vector components $F$ (\ref{eq12}) as
follows:
 \begin{equation}\label{eq14}
\begin{array}{l}
  T=\frac{\gamma-1}{\gamma R\rho_0}f_1+\frac{T_0\omega_N^2}{i\omega
  g}f_2,\\
  v_x=\frac{k_x}{\omega}f_1,\\
  \rho=\frac{1}{\gamma RT_0}f_1-\frac{\rho_0\omega_N^2}{i\omega
  g}f_2,\\
  T'=\frac{1}{R\rho_0}f'_1-\frac{T'_0}{\rho_0}\rho-\frac{\rho'_0}{\rho_0}\rho-\frac{T_0}{\rho_0}\rho'.
  \end{array}
\end{equation}
In the last formula,  $f'_1$ and $\rho'$  are obtained from Eq.
(\ref{eq3}), (\ref{eq12}), (\ref{eq13}). Thus, Eqs. (\ref{eq3}),
(\ref{eq12})-(\ref{eq14}) describe wave disturbances in the case
of dissipationless atmosphere $(n=2)$.
\subsection{Weakly dissipative atmosphere approximation, $n=2$ (Case III-a)}\label{section2.4}
In the region of the atmosphere where dissipation is small, it is
possible to use the dissipationless approximation. But it is
clear, if we can take dissipation into account, that accuracy of a
solution will be higher. It is easy to show that in the case of
small dissipation the equations with dissipation can be reduced to
a second-order problem, if we exclude from consideration too
small-scale waves.  For these waves, equations can be obtained
from (\ref{eq7}) with use of  dissipationless relations when
deriving of the dissipative terms. In this manner, as in
(\ref{eq3}), (\ref{eq13}) of Case III, we get a set of two
first-order differential equations, but it consider small
dissipation. In more detail, realization of this method is as
follows. From (\ref{eq7}), we get the relations:
 \begin{equation}\label{eq56}
\begin{array}{l}
  p'=a_{11}p+a_{12}v_z+iHR\rho_0s(T''-k_x^2T),\\
v_z'=a_{21}p+a_{22}v_z+\frac{\omega H^2}{T_0}s(T''-k_x^2T)
  \end{array}
\end{equation}
Further for  $T$ and $T''$ , we obtain the expressions in the form
of linear combinations $p,v_z$  from dissipative equations
(\ref{eq11}) We have the following from the equation of state:
 \begin{equation}\label{eq57}
\begin{array}{l}
 T=\frac{1}{R\rho_0}p-\frac{T_0}{\rho_0}\rho,\\
 T''=\frac{1}{R\rho_0}p''-\frac{T_0}{\rho_0}\rho''+2\left(\frac{1}{R\rho_0}\right)^\prime
 p'-2\left(\frac{T_0}{\rho_0}\right)^\prime
 \rho'+\left(\frac{1}{R\rho_0}\right)^{\prime\prime}p-\left(\frac{T_0}{\rho_0}\right)^{\prime\prime}
 \rho.
  \end{array}
\end{equation}
We express $\rho,\rho'$  and $\rho''$  from the fourth equation of
(\ref{eq11}):
 \begin{equation}\label{eq58}
\begin{array}{l}
 \rho=\frac{i\omega}{g}\rho_0v_z-\frac{1}{g}p'\\
\rho'=\frac{i\omega}{g}\rho_0v'_z+\frac{i\omega}{g}\rho'_0v_z-\frac{1}{g}p'',\\
\rho''=\frac{i\omega}{g}\rho_0v''_z+\frac{2i\omega}{g}\rho'_0v'_z+\frac{i\omega}{g}\rho''_0v_z-\frac{1}{g}p'''.
  \end{array}
\end{equation}
Then $p',p'',p''',v'_z,v''_z$  from (\ref{eq58}) are expressed
from (\ref{eq3}), (\ref{eq11}), (\ref{eq12}):
 \begin{equation}\label{eq59}
\begin{array}{l}
  p'=a_{11}p+a_{12}v_z,\\
v_z'=a_{21}p+a_{22}v_z,\\
  p''=a_{11}p'+a_{12}v'_z+a'_{11}p+a'_{12}v_z,\\
v_z''=a_{21}p'+a_{22}v'_z+a'_{21}p+a'_{22}v_z,\\
  p'''=a_{11}p''+a_{12}v''_z+2a'_{11}p'+2a'_{12}v'_z+a''_{11}p+a''_{12}v_z.
  \end{array}
\end{equation}
Eqs. (\ref{eq57})-(\ref{eq59}) allow us to reduce system
(\ref{eq56}) to the normal form (\ref{eq3}):
 \begin{equation}\label{eq60}
\left(^p_{v_z}\right)^\prime=\tilde{A}\left(^p_{v_z}\right)=\left(^{\tilde{a}_{11}
\tilde{a}_{12}}_{\tilde{a}_{21}
\tilde{a}_{22}}\right)\left(^p_{v_z}\right).
\end{equation}
The solutions of this system of equations give only ordinary
atmospheric waves under small dissipation effect. They do not
contain purely "dissipation" (small-scale) solutions intrinsic to
the input system of dissipative wave equations with higher order
of the differential equations.
\section{Analysis of dissipation influence on the wave propagation in isothermal atmosphere}\label{section3}
The model of an isothermal dissipative atmosphere is required for
construction DSAS to have the part of DSAS in the isothermal upper
atmosphere, besides it allows  to see specific properties of wave
propagation in the upper atmosphere, with dissipation effect
exponentially growing with height.  These properties are intrinsic
not only to an isothermal model atmosphere, but also to the real
upper atmosphere; their analysis is required for correct matching
the parts of DSAS in the upper atmosphere and lower atmosphere.
The circumstance that it is possible to reduce the wave problem to
one differential equation, which is sixth-order in the most
general case , is significant for  successful analysis of
properties of the wave propagation in the case of isothermal
dissipative atmosphere.

Consider an isothermal atmosphere with constant coefficients of
thermal conductivity and viscosity, which is formally determined
in the whole space $z\in[-\infty ,\infty]$:
\begin{equation}\label{eq15}
\begin{array}{l}
  p_0(z)=p_0(z_r)e^{-\frac{z-z_{r}}{H}}, \\
  \rho_0(z)=\rho_0(z_r)e^{-\frac{z-z_{r}}{H}}, \\
  H=\frac{RT_0}{g}, \kappa=const, \nu_1=const, \nu_2=const.
\end{array}
\end{equation}
Here $H$ is the scale height of the atmosphere;  $z_r$ is the
reference height with specified undisturbed pressure and density.
For the convenience of the wave description, we will exploit a
completely dimensionless form of its representation, i.e.
coordinates, time, wave parameters, and disturbance function will
be represented by corresponding dimensionless values:
 \begin{equation}\label{eq16}
\begin{array}{l}
  {\bf r}^*\equiv(x^*,y^*,z^*)={\bf r}/H\equiv(x,y,z)/H, t^*=t\sqrt{g/H},  \\
  k=k_xH, \sigma=\omega\sqrt{H/g}, \\
  n=\rho'/\rho_0, f=p'/p_0, \Theta=T'/T_0, \\
  u=v_x/c_s, w=v_y/c_s, v=v_z/c_s.
\end{array}
\end{equation}
To bring the disturbance equations to dimensionless form, we will
use dimensionless expressions of kinematic dissipative values:
\begin{equation}\label{eq17}
\begin{array}{l}
  s(z)=\frac{\kappa}{\sigma\gamma c_vH\sqrt{gh}}\rho_0^{-1}, \\
  \mu(z)=\frac{\nu_1}{\sigma H\sqrt{gh}}\rho_0^{-1}, \\
  q(z)=\frac{\nu_1/3+\nu_2}{\sigma H\sqrt{gh}}\rho_0^{-1}.
\end{array}
\end{equation}
By applying introduced determinations in Eqs.
(\ref{eq15})-(\ref{eq17}) to system of equations (\ref{eq1}), we
obtain complete set of equations for disturbances of density,
pressure, velocity, and temperature in the form:
\begin{equation}\label{eq18}
\begin{array}{l}
  a) \ \ \ \ \Theta=f-n, \\
  b) \ \ \ \ -i\sigma n+\Psi-\sqrt{\gamma}v=0, \\
  c) \ \ \ \ -i\sigma f-\sqrt{\gamma}v+\gamma\Psi=\sigma\gamma s\Delta^*\Theta, \\
  d) \ \ \ \ -i\sigma \sqrt{\gamma}u+ikf=\sqrt{\gamma}\sigma\mu\Delta^* u+ik\sigma q\Psi, \\
  e) \ \ \ \ -i\sigma \sqrt{\gamma}v+\dot{f}-f+n=\sqrt{\gamma}\sigma\mu\Delta^* v+\sigma q\dot{\Psi}, \\
  f) \ \ \ \ -i\sigma w=\sigma\mu\Delta^* w.
  \end{array}
\end{equation}
Here the dot is the derivative of a function with respect to $z^*$
 argument;$\Psi=\sqrt{\gamma}(\dot{v}+iku)$
 is the dimensionless divergence of velocity disturbance;
 $\Delta^*=\frac{d^2}{dz^{*2}}-k^2$is the dimensionless Laplacian.
 Eq. (\ref{eq18}f) describes an independent viscous solution unrelated to the disturbances we are interested in.
 Thus, from now on we set $w=0$  and will consider the system of five Eqs. (\ref{eq18}a)-(\ref{eq18}e) with unknowns  $(\Theta,n,f,u,v)$.
 The kinematic dissipative coefficients $s,\mu$ and $q$ are functions which grow exponentially with height.
 At low heights, where these coefficients may be ignored, set of equations (\ref{eq18}) describes ordinary classical
 acoustic and gravitational oscillations.
 In the real atmosphere, thermal conductivity exceeds viscosity. Hence with an increase in height thermal-conductivity dissipation does occur first,
 where the dimensionless function $s$ becomes of order of unity. A corresponding height in the real atmosphere may be determined, using first
 formula of Eqs. (\ref{eq17}), from the  condition
 \begin{equation}\label{eq19}
\frac{\kappa}{\omega\gamma c_v\left(H(z_c)\right)^2\rho_0(z_c)}=1.
\end{equation}
Here the height dependence of the scale height of the atmosphere
is taken into account. The value of $z_c$  may be found by solving
implicit Eq. (\ref{eq19}). Reference to Eq. (\ref{eq19}) shows
that with an increase in frequency of oscillations the critical
height should also increase. Next, we will assume for convenience
that the point of reference of the dimensionless coordinate $z^*$
corresponds to $z_c$:
 \begin{equation}\label{eq20}
z^*=(z-z_c)/H.
\end{equation}
Accordingly, $p_0(z_r)$ and $\rho_0(z_r)$ in Eq. (\ref{eq15}) are
determined by values at $z_r=z_c$ and, therefore, are implicit
functions of frequency of oscillations too. From Eqs. (\ref{eq19})
and (\ref{eq20}), expressions for $s,\mu$, and $q$ take the
following form:
\begin{equation}\label{eq21}
\begin{array}{l}
  s=e^{z^*}, \\
  \mu=\frac{\nu_1}{\kappa}c_ve^{z^*}, \\
  q=\frac{\nu_1/3+\nu_2}{\kappa}c_ve^{z^*}.
\end{array}
\end{equation}
System (\ref{eq18}a)-(\ref{eq18}e) allows reducing to one
sixth-order ordinary differential equation with one variable
$\Theta$. To derive this equation, we exclude variables in the
following sequence. First, we exclude variables $n$ and $f$ from
Eqs. (\ref{eq18}a), (\ref{eq18}b):
  \begin{equation}\label{eq22}
n=\frac{1}{i\sigma}(\Psi-\sqrt{\gamma}v),  \ \ \
f=\Theta+n=\Theta+\frac{1}{i\sigma}(\Psi-\sqrt{\gamma}v)
\end{equation}
Using (\ref{eq22}), we obtain the following expressions:
\begin{equation}\label{eq23}
\begin{array}{l}
  a) \ \ \ \ \Psi=\frac{\sigma}{\gamma-1}(\gamma
  s\Delta^*\Theta+i\Theta),\\
  b) \ \ \ \ -i\sigma \sqrt{\gamma}u+\frac{k}{\sigma}\sqrt{\gamma}v+\sqrt{\gamma}\sigma\mu\Delta^* u=ik\Theta-ik\left(\sigma q-\frac{1}{i\sigma}\right)\Psi, \\
  c) \ \ \ \ -i\sigma \sqrt{\gamma}v+\frac{1}{i\sigma}\sqrt{\gamma}\dot v+\sqrt{\gamma}\sigma\mu\Delta^* v=\dot\Theta-\Theta-\left(\sigma q-\frac{1}{i\sigma}\right)\dot\Psi, \\
  d) \ \ \ \ \Psi=\sqrt{\gamma}(\dot{v}+iku), \\
  e) \ \ \ \ \theta=\sqrt{\gamma}(\dot u-ikv), \\
  f) \ \ \ \ \Delta^* u=(\dot\theta+ik\Psi)/\sqrt{\gamma}, \\
  g) \ \ \ \ \Delta^* v=(\dot\Psi+ik\theta)/\sqrt{\gamma}.
\end{array}
\end{equation}
Here Eq. (\ref{eq23}a) results from the subtraction of Eq.
(\ref{eq18}b) from Eq. (\ref{eq18}c) with the use of Eq.
(\ref{eq18}a); Eqs. (\ref{eq23}b) and (\ref{eq23}c) result from
the substitution of Eq. (\ref{eq22}) into Eq. (\ref{eq18}d) and
Eq. (\ref{eq18}e) respectively; Eq. (\ref{eq23}d) gives the
divergence; Eq. (\ref{eq23}e) is a new auxiliary function of
current; Eq. (\ref{eq23}f) and Eq. (\ref{eq23}g) are auxiliary
identities evident from the determinations of divergence and
current function. Eq. (\ref{eq23}a) gives us the divergence
through temperature  $\Theta$. The current function may also be
expressed through $\Theta$ by summing up differentiated Eq.
(\ref{eq23}c) and Eq. (\ref{eq23}b) multiplied by $ik$:
   \begin{equation}\label{eq24}
\theta=\frac{\sigma}{k}\frac{1}{1+i\sigma^2\mu}\left\{
\hat{L}\left[\sigma(\mu+q)\Psi-\Theta-\frac{1}{i\sigma}\Psi\right]+i\sigma\Psi\right\},
\end{equation}
where: $\hat{L}=\Delta^*-\frac{d}{dz^*}$. By differentiating Eq.
\ref{eq23}b) and subtracting Eq. \ref{eq23}c) multiplied by $ik$,
we obtain the differential equation expressed through one unknown
function $\Theta$:
    \begin{equation}\label{eq25}
\left(1-i\hat{L}\mu\right)\theta+k(\mu+q)\Psi-\frac{k}{\sigma}\Theta=0,
\end{equation}
Or in more detail
   \begin{equation}\label{eq26}
\left(1-i\hat{L}\mu\right)\frac{\hat{L}\left[\sigma(\mu+q)\Psi-\Theta-\frac{1}{i\sigma}\Psi\right]+i\sigma\Psi}{1+i\sigma^2\mu}+\frac{k^2}{\sigma}(\mu+q)\Psi-\frac{k^2}{\sigma^2}\Theta=0.
\end{equation}
Other values are expressed via $\Theta$  and $\Psi$:
        \begin{equation}\label{eq27}
\begin{array}{l}
  v=\frac{1}{\sqrt\gamma}\frac{\sigma^2}{\sigma^4-k^2}\times \\
   \left[\left(1-\frac{d}{dz^*}-\frac{k^2}{\sigma^2}\right)(\Psi+i\sigma\Theta)+i\sigma^2(\mu+q)\left(\frac{d}{dz^*}+\frac{k^2}{\sigma^2}\right)\Psi-k\mu\left(\frac{d}{dz^*}+\sigma^2\right)\theta   \right], \\
  f=\Theta+\frac{i}{\sigma}(\sqrt{\gamma}v-\Psi), \\
  u=\left(\frac{k}{\sigma}f+i\mu\dot{\theta}-k(\mu+q)\Psi\right)\frac{1}{\sqrt\gamma},\\
  n=f-\Theta
\end{array}
\end{equation}
Eq. (\ref{eq26}) and Eqs. (\ref{eq27}) completely describe the
wave disturbance in the chosen approximation. Eq. (\ref{eq26})
does not have analytical solutions, but it may be used for
analyzing the asymptotic behavior of solutions at large and small
$z^*$, or for numerical solution (see \cite{Rudenko_a,Rudenko_b})
Unlike the classical dissipationless solution, Eq. (\ref{eq26})
allows the solution without an infinite increase in amplitude of
relative values of disturbances; i.e. in the whole space, the
solution may satisfy the  linear approximation \citep{Rudenko_a}.
\subsection{AGW  solution for the heat-conducting isothermal atmosphere}\label{section3.1} The most
interesting possibility of deriving an analytical form of
dissipative solutions which describes disturbances of acoustic and
gravitational ranges is provided by the approximation of
non-viscous heat-conducting atmosphere $(\mu = q = 0 )$.  In this
case, Eqs. (\ref{eq26}) and (\ref{eq27}) becomes:
   \begin{equation}\label{eq28}
\left[-\hat{L}\left(\Theta+\frac{1}{i\sigma}\Psi\right)+i\sigma\Psi\right]-\frac{k^2}{\sigma^2}\Theta=0;
\end{equation}
\begin{equation}\label{eq29}
\begin{array}{l}
  \Psi=\frac{\sigma}{\gamma-1}\left(\gamma e^{z^*}\Delta^*\Theta+i\Theta\right), \\
  v=\frac{1}{\sqrt\gamma}\frac{\sigma^2}{\sigma^4-k^2}\left(1-\frac{d}{dz^*}-\frac{k^2}{\sigma^2}\right)(\Psi+i\sigma\Theta), \\
  f=\Theta+\frac{i}{\sigma}(\sqrt\gamma v-\Psi),\\
  u=\frac{k}{\sqrt\gamma\sigma}f, \\
  n=f-\Theta.
\end{array}
\end{equation}

Introducing a new variable
    \begin{equation}\label{eq30}
\xi=\exp\left(-z^*+i\pi\frac{3}{2}\right)
\end{equation}
allows us to represent Eq. (\ref{eq28}) in the canonical form of
the generalized hypergeometric equation
   \begin{equation}\label{eq31}
\left[\xi\prod^2_{j=1}(\delta-a_j+1)-\prod^4_{i=1}(\delta-b_i)\right]\Theta=0,
\end{equation}
where  $\delta=\xi d/d\xi$, $a_{1,2}=\frac{1}{2}\pm iq$,
 $b_{1,2}=\pm k$, $b_{3,4}=1/2\pm\alpha$
 \begin{equation}\label{eq32}
\begin{array}{l}
  q=\sqrt{-\frac{1}{4}+\frac{\gamma-1}{\gamma}\frac{k^2}{\sigma^2}+\frac{\sigma^2}{\gamma}-k^2}, \\
  \alpha=\sqrt{\frac{1}{4}+k^2-\sigma^2}.

\end{array}
\end{equation}
Eq. (\ref{eq31}) has two singular points:  $\xi=0$  (the regular
singular point,  $z^*=+\infty$) and  $\xi=\infty$   (the irregular
singular point, $z^*=-\infty$)). The fundamental system of
solutions of Eq. (\ref{eq31}) may be expressed by four linearly
independent generalized Meijer functions (\cite{Luke}):
 \begin{equation}\label{eq33}
\begin{array}{l}
  a) \ \ \ \ \ \Theta_1=G^{4,1}_{2,4}\left(\xi e^{-i\pi}\left|^{a_1,a_2}_{b_1,b_2,b_3,b_4}\right.\right),\\
  b) \ \ \ \ \ \Theta_2=G^{4,1}_{2,4}\left(\xi e^{-i\pi}\left|^{a_2,a_1}_{b_1,b_2,b_3,b_4}\right.\right),\\
  c) \ \ \ \ \ \Theta_3=G^{4,0}_{2,4}\left(\xi e^{-2i\pi}\left|^{a_1,a_2}_{b_1,b_2,b_3,b_4}\right.\right), \\
  d) \ \ \ \ \ \Theta_4=G^{4,0}_{2,4}\left(\xi\left|^{a_1,a_2}_{b_1,b_2,b_3,b_4}\right.\right).
  \end{array}
\end{equation}
Desired meaningful solution of Eq. (\ref{eq28}) may be found with
use of the known properties of asymptotic behaviors of
$\Theta_i$-functions near two singular points in Eq. (\ref{eq31}).
Near the irregular point $\xi=\infty$ ($z^*=-\infty$), we have:
 \begin{equation}\label{eq34}
\begin{array}{l}
  a,b) \ \ \  \Theta_{1,2}(\xi)\sim\Theta_{1,2}^{\infty}(\xi)=p_{1,2}\cdot e^{(1/2\mp iq)z^*}=\\
  \ \ \ \frac{\Gamma\left(\frac{1}{2}\mp iq+k\right)\Gamma\left(\frac{1}{2}\mp iq-k\right)\Gamma(1\mp iq+\alpha)\Gamma(1\mp iq-\alpha)}
  {\Gamma(1\mp 2iq)}e^{\mp\frac{\pi q}{2}-i\frac{\pi}{4}}\cdot e^{(1/2\mp iq)z^*},\\
  \\
  c,d) \ \ \  \Theta_{3,4}(\xi)\sim\Theta_{3,4}^{\infty}(\xi)=\pi^{1/2}e^{-i\frac{\pi}{8}(1\mp 2)}\cdot e^{z^*/4}e^{\mp\sqrt{2}(1-i)e^{-z^*/2}}.

  \end{array}
\end{equation}
From Eqs. (\ref{eq34}a), (\ref{eq34}b) follows that, at real $q$,
asymptotics of $\Theta_1$ and  $\Theta_2$ correspond to two
different classical dissipationless waves. Here we will assume
that $\Theta_1$ corresponds to an upward propagating wave (from
the Earth); $\Theta_2$, to a downward propagating wave (towards
the Earth.) Then, for $\sigma$ and $k$ corresponding to acoustic
waves, we set $q < 0$; for those corresponding to internal gravity
waves, $q> 0$. Asymptotics of $\Theta_3$ and  $\Theta_4$ specified
by Eqs. (\ref{eq34}c), (\ref{eq34}d) correspond to dissipative
oscillations. One of these asymptotics is function of extremely
rapid decrease and the other is function of extremely rapid
increase. Decreasing $\Theta_3$ leaves the physical scene very
quickly when $z^*$ decreases; and a coefficient of the solution
with $\Theta_4$ asymptotic behavior must be chosen equal to $0$.

The behavior of the solutions  $\Theta_1$,  $\Theta_2$, and
$\Theta_3$ nearby the regular point $\xi =0(z^*=+\infty )$ is
represented by the following expressions:
    \begin{equation}\label{eq35}
\Theta_i\sim\Theta_i^0=\sum^4_{j=1}t_{ij}e^{-b_jz^*} \ \ \
(i=1,2,3).
\end{equation}
Expressions for $t_{ij}$ are derived from known asymptotics of the
Meijer G-function at the regular singular point (\cite{Luke}).
 Their explicit expressions are listed in \ref{A}.

When constructing the meaningful solution, we will assume that
there should not be upward growing asymptotic terms $\sim
e^{-b_2z^*}$ and $\sim e^{-b_4z^*}$ (from the Earth).  Besides, we
set the incident wave amplitude $= 1$. Accordingly, the desired
solution is found in the following form
    \begin{equation}\label{eq36}
\Theta(\xi)=p_1^{-1}\Theta_1(\xi)+\alpha_2\Theta_2(\xi)+\alpha_3\Theta_3(\xi),
\end{equation}
where coefficients $\alpha_2$ and $\alpha_3$ are chosen to satisfy
the condition of elimination of growing asymptotics near the
regular point of Eq. (\ref{eq31}):
 \begin{equation}\label{eq37}
\begin{array}{l}
 p_1^{-1}t_{12}+\alpha_2t_{22}+\alpha_3t_{32}=0 ,\\
  p_1^{-1}t_{13}+\alpha_2t_{23}+\alpha_3t_{33}=0 .
  \end{array}
\end{equation}
Solving (\ref{eq37}) yields:
 \begin{equation}\label{eq38}
\begin{array}{l}
  a) \ \ \ \ \  \alpha_2=-p_1^{-1}e^{-2\pi q}\frac{\sin[\pi(\alpha-iq)]\cos[\pi(k-iq)]}{\sin[\pi(\alpha+iq)]\cos[\pi(k+iq)]},\\
  b) \ \ \ \ \  \alpha_3=2\pi p_1^{-1}\frac{e^{-\pi q}}{e^{i2\pi\alpha}+e^{i2\pi k}}\left\{\frac{\sin[\pi(\alpha-iq)]}{\sin[\pi(\alpha+iq)]}- \frac{\cos[\pi(k-iq)]}{\cos[\pi(k+iq)]}\right\}.

  \end{array}
\end{equation}
Eqs. (\ref{eq36}) and (\ref{eq38}a), ((\ref{eq38}b) give the
analytical expression of the desired meaningful solution. For
$|\xi|<1$ ($z^*>0$), solution of Eq. (\ref{eq36}) can be expressed
through generalized hypergeometric functions $_mF_n$ which, in the
region of such values of arguments, are represented by simple
convergent power series, suitable for the numerical calculation.
Such a representation of solution is obtained from the standard
representation of the Meijer  $G$-function.
  \begin{equation}\label{eq39}
\begin{array}{l}
  G^{m,n}_{p,q}\left(y\left|^{a_p}_{b_q}\right.\right)=\sum^m_{h=1}\frac{\prod^m_{j=1}\Gamma(b_j-b_h)^*\prod^n_{j=1}\Gamma(1-a_j-b_h)}{\prod^q_{j=m+1}\Gamma(1-b_h-b_j)\prod^n_{j=n+1}\Gamma(a_j-b_h)}y^{b_h}\times\\
  \ \ \  _pF_{q-1}\left(\left.^{1+b_h-a_p}_{1+b_h-b_q^*}\right|(-1)^{p-m-n}y\right).
  \end{array}
\end{equation}
Here (*) implies that a term with an index equal to $h$ is
omitted. With Eq. (\ref{eq39}), after the rather lengthy
calculations, we can reduce solution of Eq. (\ref{eq36}) to the
following form:
  \begin{equation}\label{eq40}
\begin{array}{l}
\Theta(z^*>0)=\beta_0\beta_1 e^{-kz^*}\times _2F_3\left(\left.^{\frac{1}{2}+k-iq,\frac{1}{2}+k+iq}_{1+2k,\frac{1}{2}+k+\alpha,\frac{1}{2}+k-\alpha}\right|-ie^{-z^*}\right)+ \\
 \ \ \ \beta_0\beta_2 e^{-(1/2+\alpha)z^*}\times _2F_3\left(\left.^{1+\alpha-iq,1+\alpha+iq}_{\frac{3}{2}+\alpha-k,\frac{3}{2}+\alpha+k,1+2\alpha}\right|-ie^{-z^*}\right),
  \end{array}
\end{equation}
where \\
 $\beta_0=\frac{\Gamma\left(\frac{1}{2}+iq+k\right)\Gamma(iq+\alpha)}{\Gamma(2iq)\Gamma\left(\frac{1}{2}+\alpha+k\right)}e^{-i\frac{\pi}{4}-\pi\frac{q}{2}}$,
$\beta_1=\frac{\Gamma\left(\frac{1}{2}+\alpha-k\right)\Gamma\left(\frac{1}{2}+iq+k\right)}{\Gamma(1+2k)\Gamma(1-iq+\alpha)}e^{-i\frac{\pi}{2}k}$,\\
$\beta_2=\frac{\Gamma\left(-\frac{1}{2}-\alpha+k\right)\Gamma(1+iq+\alpha)}{\left(\frac{1}{2}+\alpha+k\right)\Gamma(1+2\alpha)\Gamma\left(\frac{1}{2}-iq+k\right)}e^{-i\frac{\pi}{2}\left(\frac{1}{2}+\alpha\right)}$.
\\
Given  $z^*\to+\infty$, the generalized hypergeometric functions
$F$ in Eq. (\ref{eq40}) tend to unity, and the solution  $\Theta$
takes a simple asymptotic form with two exponentially decreasing
terms.

The found solution describes the incidence of internal gravity or
acoustic wave with an arbitrary inclination to the dissipative
region  $z^*>0$, its reflection from this region, and its
penetration to this region with the transformation of it into a
dissipative form. The complex coefficient of reflection can be
expressed by a simple analytical formula:
      \begin{equation}\label{eq41}
K=\alpha_2p_2.
\end{equation}
In \cite{Rudenko_a}, the reflection coefficient module is shown to
take a value of order of unity, if the typical vertical scale of
incident wave $q^{-1}\gtrsim 1$. Otherwise, the contribution of
the reflected wave exponentially decreases with decreasing
vertical scale. This behavior of reflection is similar to the
ordinary wave reflection from an irregularity of a medium. In our
case, the scale of the irregularity is the scale  height of the
atmosphere $H$.

It is worth introducing an index characterizing the value of total
vertical dissipation index in the region ($z^*<0$):
      \begin{equation}\label{eq42}
\eta=|\Theta(0)|.
\end{equation}
This value can be calculated, using the solution of Eq.
(\ref{eq40}). In what follows, we will show that the behavior of
$\eta$, depending on a vertical wave scale, is similar to the
behavior of reflection coefficient. The value  $\eta$ also
characterizes the capability of a significant part of a wave
disturbance to penetrate to the upper region $z^*>0$. It is
reasonable that, if  $\eta$ is negligible, further wave
disturbance propagation may be neglected too.
\subsection{Classification of AGW by their dissipative properties}\label{section3.2}
The understanding of certain wave effects at heights of the upper
atmosphere requires knowing possible regimes of wave propagation
which are associated with dissipation effect. Of particular
interest first is to determine the typical height from which
dissipation assumes a dominant control over the wave process, and
the second is the type of wave propagation, depending on the
period and spatial scale of disturbance (whether it is an
approximately dissipationless propagation or propagation with
dominant dissipation.)
\subsubsection{Classification of AGW by the dissipation critical height}\label{section3.2.1}
\begin{figure}
    \includegraphics[width=1\linewidth]{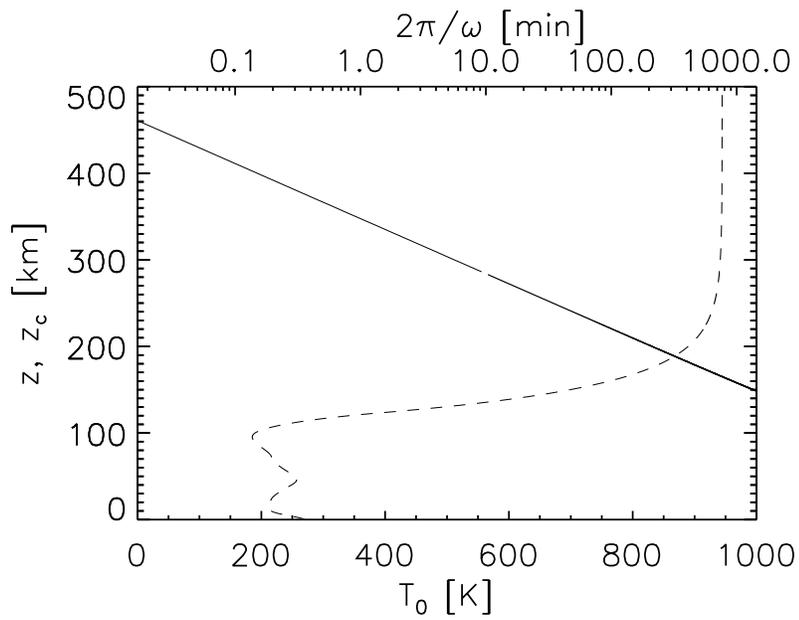}\\
  \caption{The dashed curve shows the height dependence of temperature in the selected model;
  the thick solid curve indicates the dependence of $z_c$ on the period of oscillations corresponding
  to IGWs incident from the lower atmosphere; the solid curve indicates the dependence of $z_c$ on the period of oscillations corresponding
  to acoustic waves incident from the lower atmosphere.}\label{fig1}
\end{figure}

The dependence of  $z_c$ on oscillation frequency enables a
convenient classification of waves only by their oscillation
periods. To make such a classification, we will exploit the
following model of medium: \\
-- The vertical distribution of temperature $T_0(z)$ according to
the NRLMSISE-2000 distribution with geographic coordinates of
Irkutsk for the local noon of winter opposition; \\
-- $p_0(z)=p_0(0)\exp\left[-\frac{g}{R}\int_0^z\frac{1}{T_0(z')}dz'\right], \ \ \ p_0(0)=1.01 \rm{Pa}$; \\
--  $\rho_0(z)=\rho_0(0)\exp\left[-\frac{g}{R}\int_0^z\frac{1}{T_0(z')}dz'\right], \ \ \ \rho_0(0)=287.0 \rm{g/m^3}$; \\
-- $g=9.807\rm{m/s^2}$, $R=287\rm{J/(kg\cdot K)}$,
$\kappa=0.026\rm{J/(K\cdot m\cdot s)}$,\\
$c_v=716.72\rm{J/(kg\cdot K)}$. \\
Solving Eq. (\ref{eq19}) for  $z_c$ (the plot in Figure
\ref{fig1}) provides the following useful information: a) the
height of transformation of waves of the selected period into
dissipative oscillations; b) the height above which waves of the
selected period with vertical scales much less than the height of
the atmosphere should be heavily suppressed by dissipation (in
fact, such waves should not appear at these heights); c) the
height limiting the applicability of the WKB approximation for the
wave of the selected period.
\subsubsection{Classification of AGW by their total vertical dissipation}\label{section3.2.2}
\begin{figure}
    \includegraphics[width=1\linewidth]{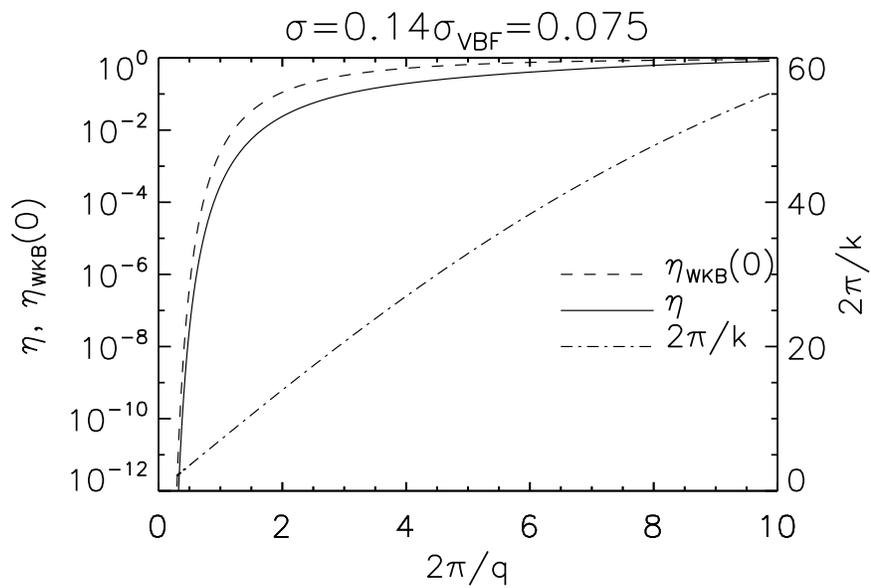}\\
  \caption{The solid line is the dependence of $\eta$ on vertical length
of the incident wave; the dashed line is the same for
$\eta_{WKB}(0)$; the dash-dotted line is the dependence of
horizontal wavelength on vertical length of the incident wave for
the selected wave solutions   (the left axis). The frequency
$\sigma$ is constant (0.14 of the Vaisala-Brent frequency
$\sigma_{VBF}$).}\label{fig2}
\end{figure}

\begin{figure}
    \includegraphics[width=1\linewidth]{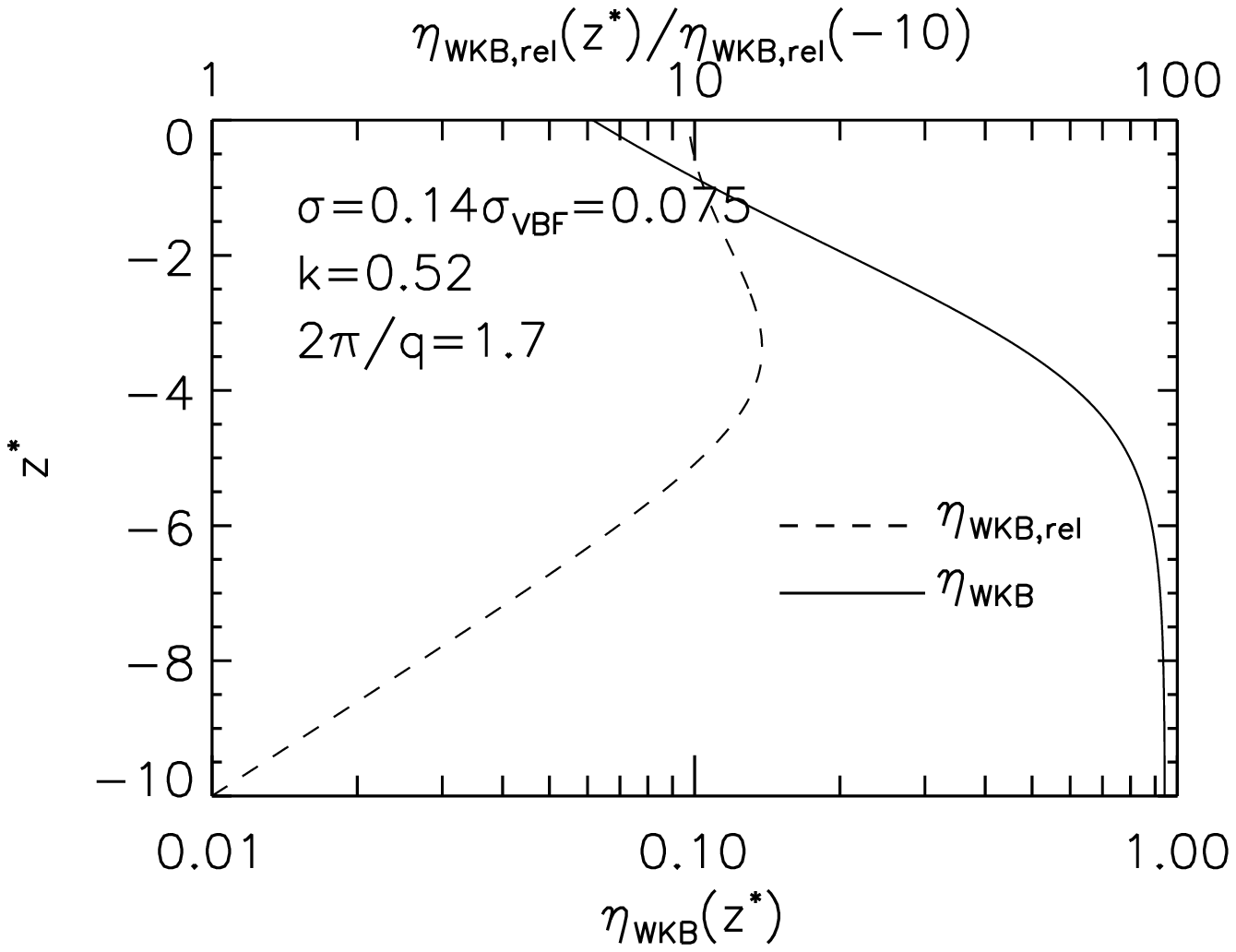}\\
  \caption{The height dependence of   $\eta_{WKB}$ and $\eta_{WKB,rel}$ at
given wave parameters.}\label{fig3}
\end{figure}

An important characteristic of a wave with certain wave parameters
$\sigma$ and $k$  is the ratio of the amplitude of the wave
solution at $z_c$ to the amplitude of the wave incident from minus
infinity  $\eta$, Eq. (\ref{eq42}). This ratio gives useful
information about the ability of this wave to have physically
meaningful values at heights of order of and higher than  $z_c$. A
direct calculation of the wave solution for the height region
$z<z_c$ from general expression of Eq. (\ref{eq36}) is unlikely
technically feasible. The first term of asymptotic expansion Eq.
(\ref{eq34}a) describes the solution only at a sufficiently large
distance from $z_c$ and does not reflect the real behavior of the
amplitude under the influence of dissipation because this term has
the form of ordinary propagating dissipationless wave. Only
beginning from  $z_c$, we can exactly calculate the solution by
power expansion Eq. (\ref{eq40}). On the other hand, the
characteristic we are interested in may be compared with its
values obtained in the WKB approximation for Eq. (\ref{eq31}). The
use of the WKB approximation also enables us to obtain the
qualitative behavior of the solution in the region  $z\lesssim
z_c$. For simplicity, consider the case when a WKB estimated value
of  $\eta$ may be expressed in an analytical form. For this
purpose, rewrite Eq. (\ref{eq31}) for a new variable m. For this
purpose, rewrite Eq. (\ref{eq14}) for a new variable
$\Theta_{ref}=\Theta e^{-\frac{1}{2}z^*}$ :
      \begin{equation}\label{eq43}
\left\{\frac{d^2}{{dz^*}^2}+q^2+\frac{e^{z^*}}{i}\left(\frac{d^2}{{dz^*}^2}-\frac{1}{4}-k^2+\sigma^2
\right)\left[\left(\frac{d}{dz^*}+\frac{1}{2}\right)^2-k^2\right]\right\}\Theta_{ref}=0.
\end{equation}
In the WKB approximation, Eq. (\ref{eq43}) yields a quartic
algebraic equation for a complex value of a dimensionless vertical
wave number   $k_z^*$:
      \begin{equation}\label{eq44}
-{k_z^*}^2+q^2+\frac{e^{z^*}}{i}\left(-{k_z^*}^2-\frac{1}{4}-k^2+\sigma^2
\right)\left[\left(ik_z^*+\frac{1}{2}\right)^2-k^2\right]=0.
\end{equation}
Note that Eq. (\ref{eq44}) can be precisely obtained from similar
dispersion Eq.  (19) from \cite{Vadas} by excluding terms
comprising first and second viscosities. For simplicity, consider
IGW continuum waves with low frequencies $\sigma$ and $k_z^*\gg
1$. In this case, Equation
       \begin{equation}\label{eq45}
-{k_z^*}^2+q^2+\frac{e^{z^*}}{i}\left({k_z^*}^2+k^2\right)^2=0
\end{equation}
gives four simple roots, one of which corresponding to upward
propagating IGW wave (from the Earth) at  $z^*\rightarrow -\infty$
will be interesting for us:
       \begin{equation}\label{eq46}
k_z^*=-\sqrt{\frac{2K^{2}}{\left( \sqrt{4ie^{z^*}K^{2}+1}+1\right)
}-k^{2}},
\end{equation}
Where  $K=\sqrt{q^2+k^2}$  is the full vector of a dissipationless
wave. The total vertical dissipation index at $z^*$  may be
presented in the form:
       \begin{equation}\label{eq47}
\eta_{WKB}=\sqrt{\frac{q}{\left|k_z^*(z^*)\right|}}e^{-{\rm
Im}\left[\int^{z^*}_{\infty}k_z^*(z^{*'})dz^{*'}\right]}=\sqrt{\frac{q}{\left|k_z^*(z^*)\right|}}e^{-\Gamma}.
\end{equation}
Expression of Eq. (\ref{eq47}) may be used for estimating the
total vertical dissipation index in the real atmosphere. In the
isothermal atmosphere approximation, the integral in the exponent
may be presented in the analytical form:
      \begin{equation}\label{eq48}
\Gamma={\rm Re}\left[i\sqrt{2}K\left(2in{\rm
Ln}\frac{a+in}{a-in}+b{\rm
Ln}\frac{b-a}{b+a}-2a-i\frac{\pi}{2\sqrt{2}}\right)\right],
\end{equation}
where $n=\frac{k}{\sqrt{2}K}$;
$a=\sqrt{\frac{1}{1+\sqrt{1+4iKe^{z^*}}}-n^2}$;
$b=\sqrt{\frac{1}{2}-n^2}$. In addition to the value determined by
Formula of Eq. (\ref{eq47}), the value of the total vertical
dissipation index versus the disturbance amplitude at the given
lower height $z_n^*$ may also be useful:
      \begin{equation}\label{eq49}
\eta_{WKB}(z^*,z_n^*)=\eta_{WKB}(z^*)/\eta_{WKB}(z_n^*).
\end{equation}
The total vertical dissipation index $\eta_{WKB}(0)$  expressed
via Eq. (\ref{eq48}) is comparable with similar value in
Eq.(\ref{eq42}) obtained from the analytical wave solution (Figure
\ref{fig2}); as function of  $z^*$, this value qualitatively
describes the amplitude of the wave solution in the isothermal
atmosphere below $z_c$ ($z^*<0$) (Figure \ref{fig3}).

Figure \ref{fig2} demonstrates convergence of the total vertical
dissipation index dependences at decreasing vertical wave scales.
This comparison shows propriety  of the analytical characteristic
of dissipation, $\eta$. The fact that the introduced
characteristic of dissipation is determined without limitations
for any wave parameters warrants its use as a universal wave
characteristic. On the other hand, the obtained convergence of
$\eta$ and  $\eta_{WKB} $ in the large-scales region may justify
using the WKB approximation for real vertically long-wave
disturbances to obtain amplitude estimates, though formally this
approximation is not applicable.

Figure \ref{fig3} illustrates the quite predictable typical
behavior of the vertical distribution of wave amplitudes below
$z_c$. It is obvious that the dissipation effect begins several
scales of the height of the atmosphere to the critical height. The
additional characteristic
 $\eta_{WKB,rel}=e^{\frac{1}{2}z^*}\eta_{WKB}$ demonstrates the required dissipation-provoked
 suppression of the exponential growth of the amplitude of relative oscillations. The suppression
 of the exponential growth provides, in turn, a possibility of satisfying the linearity of
 disturbance at all heights.
\begin{figure}
    \includegraphics[width=1\linewidth]{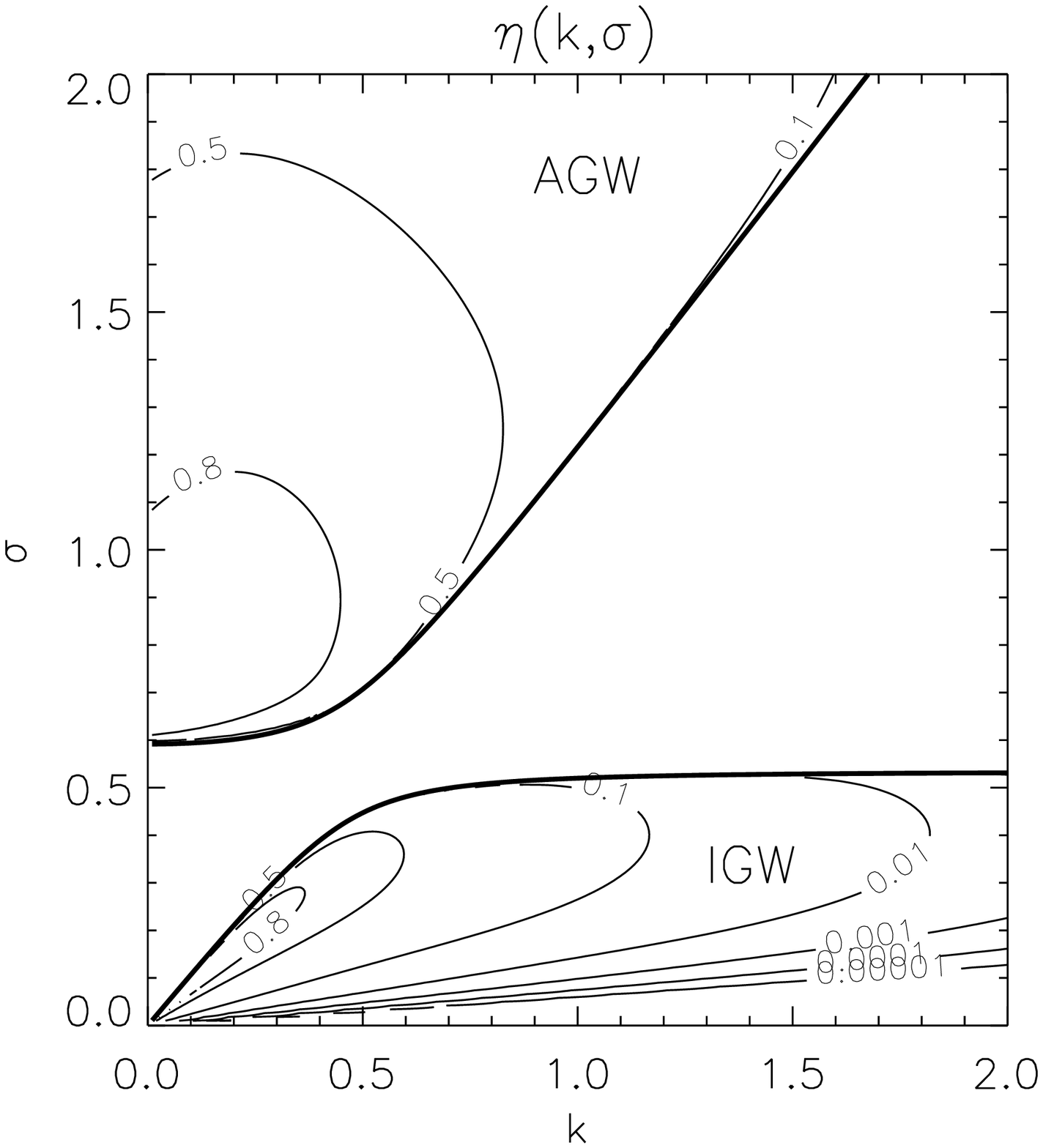}\\
  \caption{The levels of $\eta$ in the plane of $k$ and $\sigma$.}\label{fig4}
\end{figure}
\begin{figure}
    \includegraphics[width=1\linewidth]{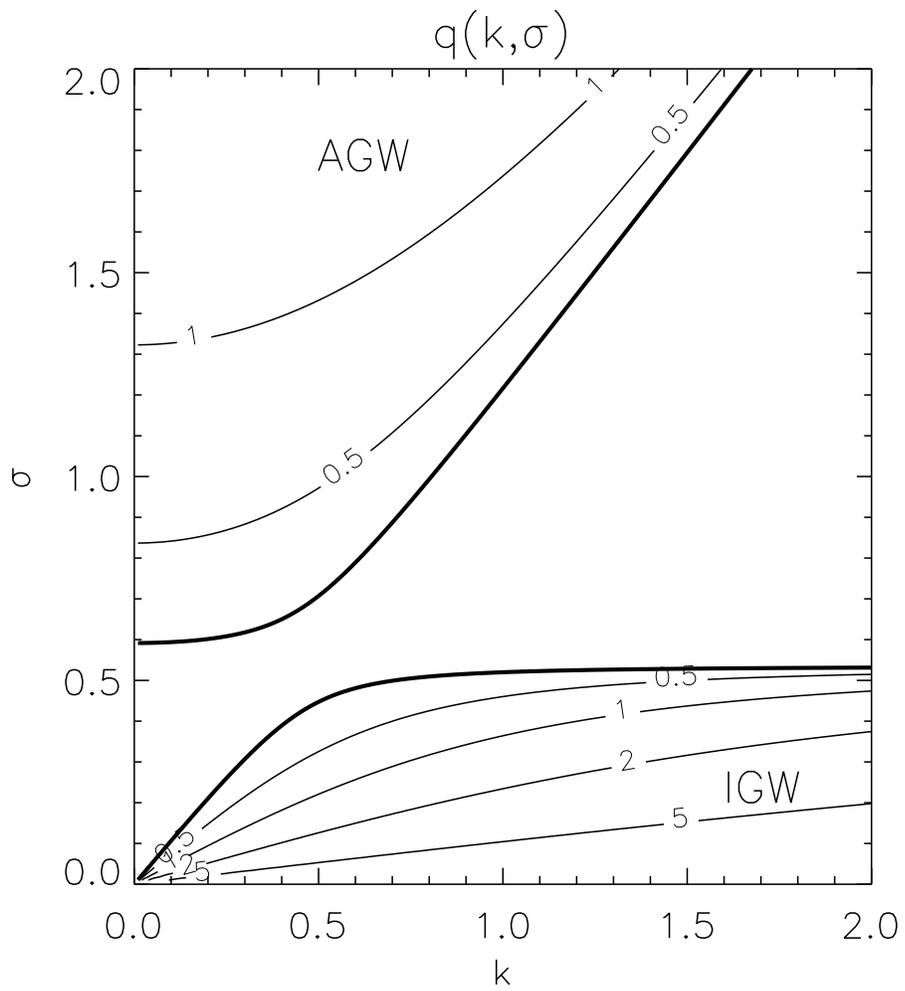}\\
  \caption{The levels of $q$ in the plane of $k$ and $\sigma$.}\label{fig5}
\end{figure}

A general picture of distribution of the total vertical
dissipation index for internal gravity and acoustic waves is given
in Figure \ref{fig4}, which shows levels of constant values $\eta$
in the plane of wave parameters $(k,\sigma )$. For convenience of
comparison, Figure \ref{fig5} presents levels of constant values
of the vertical wavenumber  $q$ in the same plane. Figure
\ref{fig4} and Figure \ref{fig5} classify waves of weak, moderate,
and strong total vertical dissipation according to  $k$, $\sigma$
and allow us to estimate the possibility of their penetration to
heights above $z_c$.
\section{Construction of DSAS for a heat-conducting non-isothermal  atmosphere}\label{section4}
The analysis in the preceding section allows us to start
constructing DSAS.  It is clear that our DSAS is bound to be a
certain combination of solutions having non-growing asymptotics
$\sim e^{-kz^*}$ and $\sim e^{-(\frac{1}{2}+\alpha)z^*}$ in the
upper atmosphere, satisfying an isothermal condition. The wave
solution found by us (Eq. (\ref{eq40}) for an everywhere
isothermal atmosphere is a sum of two terms being a subset of
fundamental solutions of Eq. (\ref{eq28}) in their turn. The sum
of these solutions does not contain the asymptotics (\ref{eq34}d)
growing extremely rapidly downward (towards the Earth) which
should not be present in a physical solution. Further we will use
the notations $..^k$ and $..^\alpha$ for any component of
disturbance with a corresponding asymptotics. Then we can write
(\ref{eq40}) in the following form:
      \begin{equation}\label{eq50}
\Theta=\Theta^{(k)}+\Theta^{(\alpha)},
\end{equation}
  \begin{equation}\label{eq51}
\begin{array}{l}
\Theta^{(k)}=\beta_0\beta_1 e^{-kz^*}\times _2F_3\left(\left.^{\frac{1}{2}+k-iq,\frac{1}{2}+k+iq}_{1+2k,\frac{1}{2}+k+\alpha,\frac{1}{2}+k-\alpha}\right|-ie^{-z^*}\right), \\
 \Theta^{(\alpha)}=\beta_0\beta_2 e^{-(1/2+\alpha)z^*}\times _2F_3\left(\left.^{1+\alpha-iq,1+\alpha+iq}_{\frac{3}{2}+\alpha-k,\frac{3}{2}+\alpha+k,1+2\alpha}\right|-ie^{-z^*}\right),
  \end{array}
\end{equation}
Our DSAS should be also presented by a sum of solutions coinciding
in the upper  isothermal part  $(I)$ of the atmosphere with
analytical solutions $..^k$ and $..^\alpha$ (Eqs. (\ref{eq51})) We
will also use $..^k$ and $..^\alpha$ to denote such solutions,
consisting of isothermal parts and their continuations into the
middle non-isothermal part $(II)$. However, it is clear that since
the downward continuations of the asymptotic    isothermal
solutions are different in the everywhere isothermal atmosphere
and in the real atmosphere, the functions  $..^k$ and $..^\alpha$
should be included in DSAS for the real atmosphere in a
combination other than (\ref{eq50}). In the case of the real
atmosphere, when a necessary combination of solutions $..^k$ and
$..^\alpha$ is unknown, we have to proceed from the consideration
that the right combination of solutions $..^k$ and $..^\alpha$  at
heights corresponding to small values of the parameter $s<<1$,
should yield, according to Formulas (\ref{eq3}),
(\ref{eq8})-(\ref{eq10}) of Case II, a solution close to a
solution of "dissipationless" problem ((\ref{eq3}),
(\ref{eq12})-(\ref{eq14}) of Case III . This condition will be the
condition of matching solutions in the upper dissipative
atmosphere and lower small-dissipation atmosphere. In its physical
sense, this condition  is equivalent to  $\Theta_4$ elimination in
the case of the everywhere isothermal atmosphere. Really, to
satisfy the matching condition, the right combination of solutions
$..^k$ and $..^\alpha$ for the real atmosphere should not contain
the rapidly growing downward solutions, like ((\ref{eq34}d), as is
the case of the solution for an everywhere isothermal atmosphere
((\ref{eq40}). As for the part of the solution for the real
atmosphere that is fast decreasing downward, like ((\ref{eq34}s),
it is negligibly small when $s<<1$, with the natural assumption
that the contribution of this part of the solution is equal to the
rest of the solution at the heights of $s\sim 1$. Having the right
combination of solutions $..^k$ and $..^\alpha$, in the upper
height range limited by a selected height with the small parameter
$s$, we get a solution satisfying both the upper boundary
condition, since  it is a combination of $..^k$ and $..^\alpha$,
and the condition of matching solutions in the upper dissipative
atmosphere and lower small-dissipation atmosphere, due to the
selection the coefficients of the combination of $..^k$ and
$..^\alpha$. Then, using the  $p$ and $v_z$ values  of the
solution for the region $s>s_t$ at a selected height $s_t$,
corresponding to a small value of the parameter $s$,as initial
conditions for continuation the solution downward, we can find a
DSAS in the lower atmosphere up to the Earth's surface.

Thus, we single out three height ranges:
  \begin{equation}\label{eq52}
\begin{array}{l}
R^I:z\in [z_I,\infty] \\
R^{II}:z\in [z_t,z_I] \\
R^{III}:z\in [0,z_t]
  \end{array}
\end{equation}
There $z_I$  is the minimum height of the isothermal range in a
considered model of the atmosphere;  $z_t$, the height
corresponding to the chosen small parameter  $s$ $(s=s_t<<1)$. In
the region $II$, we solve the Cauchy problem for the system of
equations of Case II. In this connection, we are guided by the
reversibility of the numerical solution of the Cauchy problem of
Case II in the height range $II$, when choosing   $s_t$. It is
evident that numerical instability associated with presence of
rapid asymptotic increase of the solution (similar to that in the
isothermal case (\ref{eq34}d) under too small values of the
parameter $s$ inevitably compromises the reversibility of the
Cauchy problem. The numerical research conducted by us has shown
that $0.05$  was the optimal value for $s_t$ satisfying the
condition of reversibility. Lower this threshold, the numerical
errors manifest, rapidly growing to infiniteness with the further
reduction of  $s_t$.

In the region $I$, we calculate two solutions  $..^k$ and
$..^\alpha$  by analytical Formulas (\ref{eq51}), (\ref{eq29}).
Using analytical expressions of the vectors $F$ for the solutions
$..^k$ and $..^\alpha$  at the height of $z_I$  as boundary
conditions of the wave problem of Case II, accordingly, we solve
two Cauchy problems numerically in the region $II$. After the
solutions  $..^k$ and $..^\alpha$  are obtained in the region
$I\cup II$, we start searching their right combinations. Without
loss of generality, we present the desired combination in the form
of $..^{(k)}+W..^{(\alpha)}$, where $W$ is the desired value.  In
accordance with our assumption, the values $T,T^\prime$ should
satisfy the "small-dissipation" or even "dissipationless"
equations of Case III at the height $z_t$  with good accuracy.
Designating these values as  $T_{III}=T_{III}(p,v_z)$,
$T^\prime_{III}=T^\prime_{III}(p,v_z)$ we can introduce a vector
  \begin{equation}\label{eq53}
\left(\begin{array}{c}
H(T^{(k)^\prime}-T_{III}^{(k)^\prime} )+WH(T^{(\alpha)^\prime}-T_{III}^{(\alpha)^\prime} )\\
(T^{(k)}-T_{III}^{(k)} )+WH(T^{(\alpha)}-T_{III}^{(\alpha)} )
  \end{array}\right)=
  \left(\begin{array}{c}
a_1+Wa_2 \\
b_1+Wb_2
  \end{array}\right).
\end{equation}
Minimization of the vector norm ((\ref{eq53}) as a positive
definite complex-valued bilinear form
  \begin{equation}\label{eq54}
(a_1+Wa_2)\overline{(a_1+Wa_2)}+(b_1+Wb_2)\overline{(b_1+Wb_2)}
\end{equation}
gives the following formula  for the complex coefficient $W$:
  \begin{equation}\label{eq55}
W=\frac{a_1 \overline{a_2}+b_1 \overline{b_2}}{a_2
\overline{a_2}+b_2 \overline{b_2}}.
\end{equation}
The line denotes a complex conjugate value. Formulas (\ref{eq3}),
(\ref{eq14}), (\ref{eq53}), and (\ref{eq55}) give the desired
coefficient $W$ through the values of $(p,v_z)^{(k),(\alpha)}$  at
the height $z_t$. If, obviously, the minimum value of the norm
(\ref{eq53}) is close to zero, we have smallness of the jumps of
all the components of the solution defined by Eqs. (\ref{eq14})
The calculations have showed, predictably, that matching in
accordance with our procedure is provided with precision of the
order  $s_t$. Thus, having defined  $W$, we found the desire
solution in the region $I\cup II$.

We finally complete the DSAS solving the Cauchy problem for the
system of the form  (\ref{eq60}) in the lower region $III$ ( with
use of the dissipationless or small-dissipation approximations).
The role of initial values for the solution in the region $III$ is
played by values of the solution for the region $I\cup II$  taking
at its lower boundary.
\section{Testing of DSAS  code on  everywhere isothermal atmosphere model}\label{section5}
We carried out the test calculations of DSAS for the model of an
everywhere isothermal atmosphere, using a code developed exactly
in accordance with the above described DSAS method for a
non-isothermal atmosphere. The dimensionless wave parameters
$\sigma=0.0616$, $k=0.1289$ are chosen. If a typical atmospheric
height scale is taken $H=28 \ km$, these parameters will
correspond to an oscillation with the period $T=90.84 \ min$ and
horizontal wavelength  $\lambda_{gor}=2\pi/k_x=1365 \ km$.
\subsection*{Test A:}
Figure 6 shows all wave component obtain numerically for the model
of an everywhere isothermal atmosphere. We use wave components
notations in accordance with the notations in Section
\ref{section3}.
\begin{figure}
    \includegraphics[width=1\linewidth]{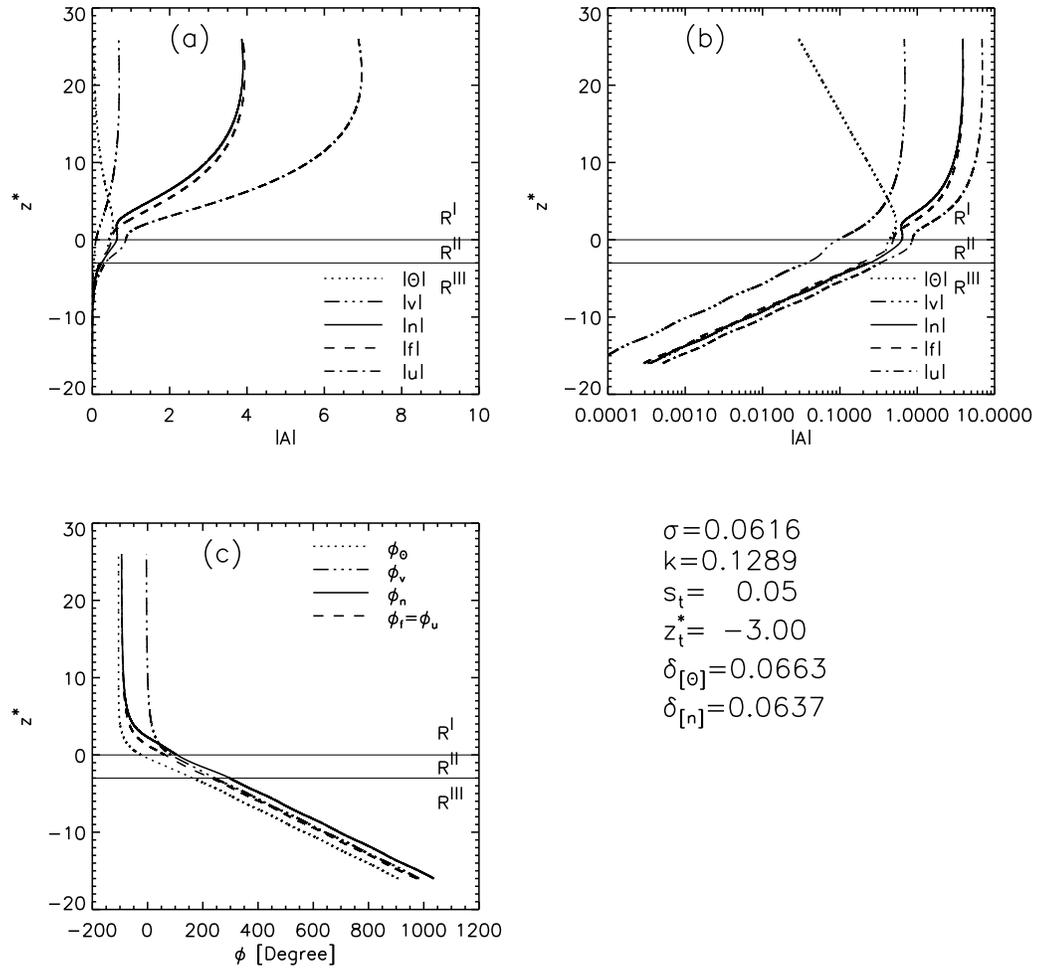}\\
  \caption{The example of a numerical DSAS for an isothermal atmosphere.}\label{fig6}
\end{figure}
The frames $(a)$ and $(b)$ of Figure \ref{fig6} show height
dependences the modules of the wave component in the normal and
logarithmic scales, accordingly. The frame $(c)$ of Figure
\ref{fig6} represents phases of the same components. In this case,
we chose the height of transfer from an analytical solution to a
numerical one equal to the critical height  $z_I^*=z_c^*$. Any
choice of $z_I^*>0$ leads to one and the same result. The
parameters $s_t$ and $z_t^*$  height corresponding to it are given
in the right lower quadrant in Figure \ref{fig6}. Solving a Cauchy
problem in the reverse order from the height $z_t^*$  of Case II,
all the height dependences are reproduced with very good acuracy
in the region  $I\cup II$.  We do not give results of these tests
because they coincide with the results of Figure \ref{fig6}.
Figure \ref{fig6} clearly convinces of sufficient smoothness of
the wave solution in all its components. The measures of the
discontinuity of the wave components $\Theta$ and $n$ at the
height $z_t^*$  are displayed in the right lower quadrant in
Figure \ref{fig6}. The notation $\delta_{[x]}$  denotes a value
  \begin{equation}\label{eq61}
\delta_{[x]}=2\frac{|X(+z_t^*)-X(-z_t^*)|}{|X(+z_t^*)+X(-z_t^*)|}.
\end{equation}
As we expected, the values of these quantities are small and have
the order of  $s_t$.
\subsection*{Test B:}
We carried out a test based on the fact that the analytical
solution of Eq. (\ref{eq40}) gives us a ready solution in the
region $I$. We used this solution for setting the Cauchy problem
in region $II$. We do not give a corresponding DSAS because it is
visually identical with what was given in Figure \ref{fig6}. But
in this case, we got some better measures of the discontinuity of
a solution:  $\delta_{[\Theta]}=0.0272$;  $\delta_{[n]}=0.0264$.
\subsection*{Test C:}
\begin{figure}
    \includegraphics[width=1\linewidth]{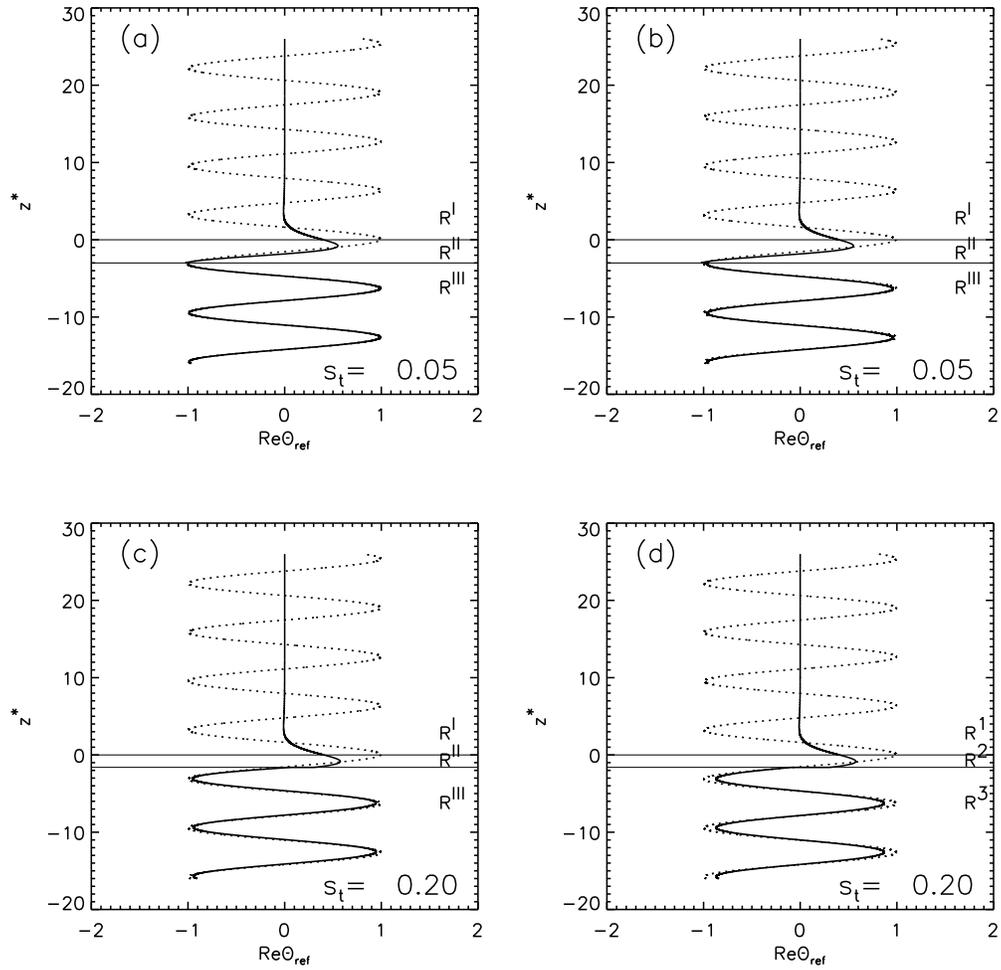}\\
  \caption{The analysis of a numerical DSAS in the lower asymptotic region.}\label{fig7}
\end{figure}

In the region $II\cup III$ we do not have the possibility of
calculating a wave solution in an analytical form because of the
difficulties caused by formal divergence of the series of the
hypergeometric functions in Eq. (\ref{eq40}). We can use only the
asymptotics of the  analytical solution at $z^*\to-\infty$  for
tests. Figure \ref{fig7} presents the comparison results of our
DSAS and the asymptotics of the analytical solution. We chose
$\texttt{Re}\Theta_{ref}=\texttt{Re}\left(\Theta
e^{-\frac{1}{2}z^*}\right)$ as a value to be tested (solid line).
For the asymptotics of the analytical solution,
$\texttt{Re}\Theta_{ref}$  is derived from Formulas
(\ref{eq34}a,b), (\ref{eq36}) and (\ref{eq41}):
$\texttt{Re}\Theta_{ref}=\texttt{Re}(e^{-iqz^*}+Ke^{iqz^*})$  (a
dashed line.) In our case, the complex coefficient $K$ is
sufficiently small, $K=-0.0055-i0.0439$.

Especially for comparison, we made calculations both taking into
account small dissipation in $III$ and without taking it into
consideration in this region. Furthermore, we, also for
comparison, used two values of $s_t$. The results are shown in
different frames of Figure \ref{fig7}.  The calculations for
frames $(a)$, $(c)$, and $(b)$, $(d)$ are distinguished by the
fact that $(a)$, $(c)$  in the region $III$ take small dissipation
into account in accordance with (\ref{eq56})-(\ref{eq60}); $(b)$,
 $(d)$, the dissipationless approximation (\ref{eq3}), (\ref{eq12})-(\ref{eq14}). Frames
$(a)$ and $(b)$ of Figure \ref{fig7} give results for the cases
with the least value  $s_t$,  $s_t=0.05$; frames $(c)$ and $(d)$,
with a large value of the parameter $s_t$,  $s_t=0.2$.  In all
cases, we clearly see that the numerical DSAS coincides with the
analytical one up phase. Comparing $(a)$ and $(b)$, one can find
only hardly noticeable advantage of result $(a)$. We can see more
clearly a positive effect of taking small dissipation into account
in region $III$ in Frame $(c)$ in comparison to $(d)$.  The
solution discontinuity indices most clearly show advantage of
taking small dissipation into account:\\
the frame $(a)$ of Figure \ref{fig7} -- $\delta_{[\Theta]}=0.0272,
\ \delta_{[n]}=0.0264$; \\
the frame $(b)$ of Figure \ref{fig7} -- $\delta_{[\Theta]}=0.1245,
\ \delta_{[n]}=0.1200$; \\
the frame $(c)$ of Figure \ref{fig7} -- $\delta_{[\Theta]}=0.3760,
\ \delta_{[n]}=0.3400$; \\
the frame $(d)$ of Figure \ref{fig7} -- $\delta_{[\Theta]}=0.4792,
\ \delta_{[n]}=0.4408$. \\
The last equalities convince us of appropriateness of  our   small
dissipation correction even at the least allowed values of the
parameter  $s_t$ and sufficiency of $0.05$  value for  $s_t$ as
well.
\section{DSAS for a non-isothermal atmosphere}\label{section6}
\begin{figure}
    \includegraphics[width=1\linewidth]{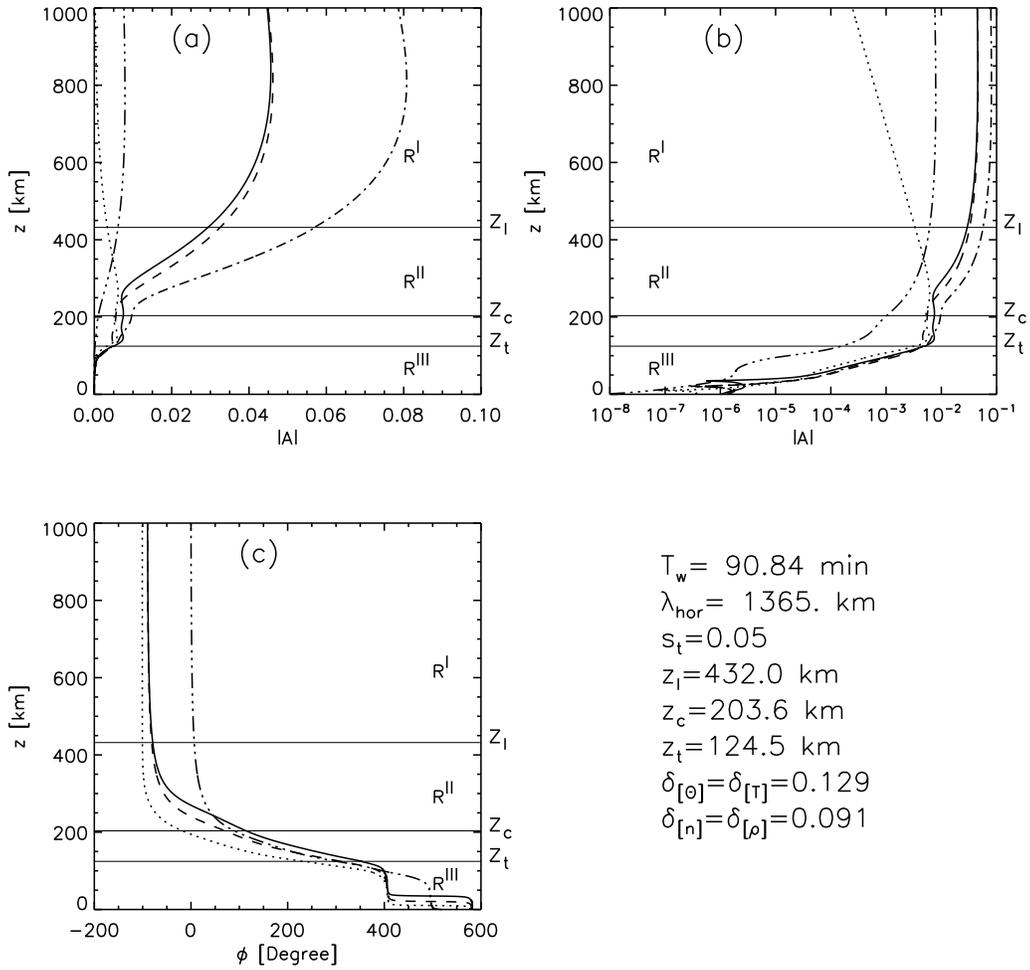}\\
  \caption{The example of a DSAS for a non-isothermal atmosphere.}\label{fig8}
\end{figure}

We used the model of the atmosphere introduced in Section
\ref{section3.2.1} with the height distribution of undisturbed
temperature shown in Figure \ref{fig1}. The calculation is carried
out for the period $T=90.84 \ min$  and horizontal wavelength
$\lambda_{hor}=1365 \ km$. The selected wave parameters $T$  and
$\lambda_{hor}$ correspond to the dimensionless $\sigma$ and $k$
used in tests in the isothermal model. The following condition is
used to norm the solution:  $\texttt{Max}(\sqrt{v_x^2+v_z^2})=50 \
m/sec$. The calculation results are shown in Figure \ref{fig8}.
They clearly convince of sufficient smoothness of DSAS in all its
components at the height of  $z_t$. The discontinuity indices
$\delta_{[\Theta]}=0.129, \ \delta_{[n]}=0.091$, approximately $5$
times higher than in an isothermal case, but remain sufficiently
small. In the upper atmosphere above $z_c$, height dependences of
the solution components are close in their nature to dependences
of the isothermal model. As in the case of the isothermal model,
we compare the received solution to the solution of the
dissipationless problem. To do so, we solve the Cauchy problem for
Eqs. (\ref{eq3}), (\ref{eq12}), (\ref{eq13}) in the reverse
direction from the Earth to the height of  $z_c$. We do not show
the received solution due to its too small graphical difference
from the solution in Figure \ref{fig8}.
\section{Green's matrix with DSAS (extended)}\label{section7}
As it is well known, the problem on evolution of disturbance from
some source can be solved without consideration of dissipation,
using the $2\times 2$ Green's matrix. However, the dissipationless
solution has the disadvantage that it cannot adequately reflect
the height structure in the upper atmospheric layers; therefore,
one should use a higher order problem than the second one there.
But one can achieve the goal of a height-structure description in
the upper atmosphere and retain the formalism of the second-order
problem to a significant extent at the same time. It is possible
to do this in that case when, as it mainly occurs in the
atmosphere, a source of disturbance is not too high, so that it
does not enter the region  of strong dissipation. For the problem
on evolution of disturbance from some source with consideration of
dissipation we will introduce the $2\times 2$ Green's modified
matrix, which we will call extended Green's matrix. Let us start
with the Green's matrix for the weakly dissipative problem.  The
inhomogeneous weakly dissipative problem can be written  in the
form
  \begin{equation}\label{eq60n}
\left(\begin{array}{c} p \\
v_z
  \end{array}\right)^\prime=
  \left(\begin{array}{cc}
  \tilde{a}_{11} & \tilde{a}_{12} \\
   \tilde{a}_{21} & \tilde{a}_{22}
  \end{array}\right)
  \left(\begin{array}{c} p \\
v_z
  \end{array}\right)
  +
  \left(\begin{array}{c} f_p \\
f_v
  \end{array}\right).
\end{equation}
The homogeneous part  of (\ref{eq60n})  is got in Section
\ref{section2}  (\ref{eq60}), the functions  $ \left(\begin{array}{c} f_p \\
f_v
  \end{array}\right)$ describe a source.
They have zero values above $z_{max}$. Eqs. (\ref{eq60n}) is a
second-order problem, and the corresponding Green's matrix has the
order $2\times 2$.   For it construction $F^u=(p^u,v_z^u)$, we
will use notations : is the weakly dissipative solution satisfying
the upper boundary condition ( the upper solution);
$F^l=(p^l,v_z^l)$, the weakly dissipative solution satisfying the
lower boundary condition ( the lower solution). For the region
$z<a$, where   $a>z_t$,  the Green's function can be written in
the form:
  \begin{equation}\label{eq62n}
  \begin{array}{l}
G_{11}=\frac{1}{W(z')}\times\left\{ \begin{array}{c}
p^u(z)v_z^l(z'), \ 0<z'<z \\
p^l(z)v_z^u(z'), \ z<z'<a
\end{array}
\right.; \\ G_{12}=\frac{1}{W(z')}\times\left\{ \begin{array}{c}
v_z^u(z)v_z^l(z'), \ 0<z'<z \\
v_z^l(z)v_z^u(z'), \ z<z'<a
\end{array}
\right.;
  \\ G_{21}=\frac{-1}{W(z')}\times\left\{ \begin{array}{c}
p^u(z)p^l(z'), \ 0<z'<z \\
p^l(z)p^u(z'), \ z<z'<a
\end{array}
\right.;  \\ G_{22}=\frac{-1}{W(z')}\times\left\{ \begin{array}{c}
v_z^u(z)p^l(z'), \ 0<z'<z \\
v_z^l(z)p^u(z'), \ z<z'<a
\end{array}
\right. .
\end{array}
\end{equation}
Here $W(z')=v_z^l()z')p^u(z')-v_z^u()z')p^l(z')$
 The  solution of (\ref{eq60n}) is
  \begin{equation}\label{eq61n}
    \begin{array}{l}
    p(z)=\int\limits_{0}^{a}G_{11}(z,z')f_p(z')dz'+\int\limits_{0}^{a}G_{21}(z,z')f_v(z')dz',
    \\
v_z(z)=\int\limits_{0}^{a}G_{12}(z,z')f_p(z')dz'+\int\limits_{0}^{a}G_{22}(z,z')f_v(z')dz'
\end{array}
\end{equation}
It is clear that, if we use DSAS as the upper solution then we
will get a solution of the dissipation problem, the homogeneous
part  of which is given by formulae (\ref{eq3}), (\ref{eq8}),
(\ref{eq9}) and the source is from (\ref{eq60n}). The disturbance
in the
upper atmospheric layers is: \\
$F(z)=F^u(z)\int\limits_{0}^{z_{max}}\frac{1}{W(z')}[v_z^l(z')f_p(z')-p^l(z')f_v(z')]dz'$
\section*{Conclusion}\label{section7}
In this paper, we proposed a method for obtaining a wave solution
above a source for the real     atmosphere (DSAS). The method of
construction of DSAS was successfully tested with the everywhere
isothermal atmosphere model.

The most essential elements of the test are: calculation of jumps
at the matching point to have demonstrated their sufficient
smallness and comparison of the numerical solution with an
asymptotics of an analytical solution in the region of small
dissipation to have demonstrated their coincidence. We have also
given the results of DSAS calculations for a real non-isothermal
atmosphere.

DSAS itself describes height structure of disturbance above any
source provided that it is in the lower part of the atmosphere. In
this paper, it is shown that one can construct an expanded Green's
matrix $2\times 2$ with use of DSAS and a solution for the lower
weekly dissipative part of the atmosphere, which, as we have
shown, can be found analogously to the dissipationless one.  The
expanded Green's matrix let us to find the disturbance produced by
some source in the all height ranges. Íerewith,  the DSAS
amplitude is completely determined.

The specific feature of DSAS is describing of dissipative tail,
formally indefinitely extended to the isothermal part of the
atmosphere. The essential circumstance is that the exponential
increase of wave values, relative to background, is changed by
their slow decrease, due to dissipation effect.  This allows the
wave to remain in the frames of the linear description, in whole
range of its existence, on condition that its amplitude in the
lower part of the atmosphere is sufficiently small. It is
important that the method proposed has prospects of the further
development in the framework of a more complex model of the
atmosphere. In particularly, including vertically stratified wind
is possible. The necessary equations for such models are written
in this paper. It is also possible to take more complete wave
dissipation into account, including not only heat conductivity but
viscosity as well.

In this case, the part of DSAS in the upper height range, where
dissipation is not small, is described, as we shown in this paper,
by the system of  six ODEs. Respectively, this system has six
independent solutions. For the asymptotic isothermal region, the
solutions taking into account both heat conductivity and viscosity
were obtain in \cite{Rudenko_b}. A combination of three such
isothermal solutions will satisfy the upper boundary condition.
The  way to ensure sufficient smoothness  of DSAS is the same as
in this paper.

 \newpage

\appendix
\section{}\label{A}
$t_{11}=\frac{\Gamma \left( -2k\right) \Gamma \left(
\frac{1}{2}+\alpha -k\right) \Gamma \left( \frac{1}{2}-\alpha
-k\right) \Gamma \left( \frac{1}{2}+k-iq\right) }{\Gamma \left(
\frac{1}{2}-iq-k\right) }e^{i\frac{\pi}{2} k}$,

$t_{12}=\frac{\Gamma \left( 2k\right) \Gamma \left(
\frac{1}{2}+\alpha +k\right) \Gamma \left( \frac{1}{2}-\alpha
+k\right) \Gamma \left( \frac{1}{2}-k-iq\right) }{\Gamma \left(
\frac{1}{2}-iq+k\right) }e^{-i\frac{\pi}{2} k}$,

$t_{13}=\frac{\Gamma \left( k-\frac{1}{2}-\alpha \right) \Gamma
\left( -k-\frac{1}{2}-\alpha \right) \Gamma \left( -2\alpha
\right) \Gamma \left( 1+\alpha -iq\right) }{\Gamma \left(
-iq-\alpha \right) }e^{i\frac{\pi}{2} \left( \frac{1}{2}+\alpha
\right) }$,

$t_{14}=\frac{\Gamma \left( k-\frac{1}{2}+\alpha \right) \Gamma
\left( -k-\frac{1}{2}+\alpha \right) \Gamma \left( 2\alpha \right)
\Gamma \left( 1-\alpha -iq\right) }{\Gamma \left( -iq+\alpha
\right) }e^{i\frac{\pi}{2} \left( \frac{1}{2}-\alpha \right) }$,

$t_{21}=\frac{\Gamma \left( -2k\right) \Gamma \left(
\frac{1}{2}+\alpha -k\right) \Gamma \left( \frac{1}{2}-\alpha
-k\right) \Gamma \left( \frac{1}{2}+k+iq\right) }{\Gamma \left(
\frac{1}{2}+iq-k\right) }e^{i\frac{\pi}{2} k}$,

$t_{22}=\frac{\Gamma \left( 2k\right) \Gamma \left(
\frac{1}{2}+\alpha +k\right) \Gamma \left( \frac{1}{2}-\alpha
+k\right) \Gamma \left( \frac{1}{2}-k+iq\right) }{\Gamma \left(
\frac{1}{2}+iq+k\right) }e^{-i\frac{\pi}{2} k}$,

$t_{23}=\frac{\Gamma \left( k-\frac{1}{2}-\alpha \right) \Gamma
\left( -k-\frac{1}{2}-\alpha \right) \Gamma \left( -2\alpha
\right) \Gamma \left( 1+\alpha +iq\right) }{\Gamma \left(
iq-\alpha \right) }e^{i\frac{\pi}{2} \left( \frac{1}{2}+\alpha
\right) }$,

$t_{24}=\frac{\Gamma \left( k-\frac{1}{2}+\alpha \right) \Gamma
\left( -k-\frac{1}{2}+\alpha \right) \Gamma \left( 2\alpha \right)
\Gamma \left( 1-\alpha +iq\right) }{\Gamma \left( iq+\alpha
\right) }e^{i\frac{\pi}{2} \left( \frac{1}{2}-\alpha \right) }$,

$t_{31}=\frac{\Gamma \left( -2k\right) \Gamma \left(
\frac{1}{2}+\alpha -k\right) \Gamma \left( \frac{1}{2}-\alpha
-k\right) }{\Gamma \left( \frac{1}{2}+iq-k\right) \Gamma \left(
\frac{1}{2}-iq-k\right) }e^{-i\frac{\pi}{2} k}$,

$t_{32}=\frac{\Gamma \left( 2k\right) \Gamma \left(
\frac{1}{2}+\alpha +k\right) \Gamma \left( \frac{1}{2}-\alpha
+k\right) }{\Gamma \left( \frac{1}{2}+iq+k\right) \Gamma \left(
\frac{1}{2}-iq+k\right) }e^{i\frac{\pi}{2} k}$,

$t_{33}=\frac{\Gamma \left( k-\frac{1}{2}-\alpha \right) \Gamma
\left( -k-\frac{1}{2}-\alpha \right) \Gamma \left( -2\alpha
\right) }{\Gamma \left( iq-\alpha \right) \Gamma \left( -iq-\alpha
\right) }e^{-i\frac{\pi}{2} \left( \frac{1}{2}+\alpha \right) }$,

$t_{34}=\frac{\Gamma \left( k-\frac{1}{2}+\alpha \right) \Gamma
\left( -k-\frac{1}{2}+\alpha \right) \Gamma \left( 2\alpha \right)
}{\Gamma \left( iq+\alpha \right) \Gamma \left( -iq+\alpha \right)
}e^{-i\frac{\pi}{2} \left( \frac{1}{2}-\alpha \right) }$.


\begin{thebibliography}{00}
\bibitem[\protect\citeauthoryear{Afraimovich at. al.}{2001}]{Afraimovich2001}
Afraimovich,~E.~L., Kosogorov,~E.~A., Lesyuta,~O.~S.,
Ushakov,~I.~I., Yakovets~Network,~A.~F. 2001. Geomagnetic control
of the spectrum of traveling ionospheric disturbances based on
data from a global GPS network. Annales Geophysicae, Volume 19,
Issue 7, 2001, pp.723-731, doi: 10.5194/angeo-19-723-2001.
\bibitem[\protect\citeauthoryear{Akmaev}{2001}]{Akmaev2001}
Akmaev,~R.~A. 2001. Simulation of large-scale dynamics in the
mesosphere and lower thermosphere with the Doppler-spread
parameterization of gravity waves: 2. Eddy mixing and the diurnal
tide. Journal of Geophysical Research: Atmospheres, Volume 106,
Issue D1, pp. 1205-1213, doi: 10.1029/2000JD900519.
\bibitem[\protect\citeauthoryear{Angelatsi and Forbes}{2002}]{Angelatsi2002}
Angelatsi~Coll,~M.; Forbes,~J.~M. 2002. Nonlinear interactions in
the upper atmosphere: The s = 1 and s = 3 nonmigrating semidiurnal
tides. Journal of Geophysical Research (Space Physics), Volume
107, Issue A8, pp. SIA 3-1, CiteID 1157, DOI 10.1029/2001JA900179.
\bibitem[\protect\citeauthoryear{Fesen}{1995}]{Fesen1995}
Fesen,~C.~G. 1995. Tidal effects on the thermosphere. Surveys in
Geophysics, Volume 13, Issue 3, pp.269-295, doi:
10.1007/BF02125771.
\bibitem[\protect\citeauthoryear{Forbes and Garrett}{1979}]{Forbes1979}
Forbes,~J.~M.; Garrett,~H.~B. 1979. Theoretical studies of
atmospheric tides. Reviews of Geophysics and Space Physics, vol.
17, Nov. 1979, p. 1951-1981, doi: 10.1029/RG017i008p01951.
\bibitem[\protect\citeauthoryear{Francis}{1973a}]{Francis1973_a}
Francis,~S.~H. 1973a. Acoustic-gravity modes and large-scale
traveling ionospheric disturbances of a realistic, dissipative
atmosphere, J.Geophys. Res., 78, 2278.
\bibitem[\protect\citeauthoryear{Francis}{1973b}]{Francis1973_b}
Francis,~S.~H. 1973b. Lower-atmospheric gravity modes and their
relation to mediumscale traveling ionospheric disturbances, J.
Geophys. Res., 78, 8289-8295.
\bibitem[\protect\citeauthoryear{Gavrilov}{1995}]{Gavrilov1995}
Gavrilov,~N.~M. 1995. Distributions of the intensity of ion
temperature perturbations in the thermosphere. Journal of
Geophysical Research, Volume 100, Issue A12, p. 23835-23844, doi:
10.1029/95JA01927, 1995.
\bibitem[\protect\citeauthoryear{Gavrilov, and Kshevetskii}{2014}]{Gavrilov2014}
Gavrilov,~N.~M.; Kshevetskii,~S.~P. 2014. Numerical modeling of
the propagation of nonlinear acoustic-gravity waves in the middle
and upper atmosphere. Izvestiya, Atmospheric and Oceanic Physics,
Volume 50, Issue 1, pp.66-72.
\bibitem[\protect\citeauthoryear{Gossard and Hooke}{1975}]{Hook}
Gossard,~E.~E. and Hooke,~W.~H. 1975. Waves in the Atmosphere,
Elsevier Scientific Publishing Company, New York, 456 pp.
\bibitem[\protect\citeauthoryear{Grigor'ev}{1999}]{Grigorev1999}
Grigor'ev,~G.~I. 1999. Acoustic-gravity waves in the earth's
atmosphere (review). Radiophysics and Quantum Electronics, Volume
42, Issue 1, pp.1-21, doi: 10.1007/BF02677636.
\bibitem[\protect\citeauthoryear{Heale at. al.}{2014}]{Heale2014}
Heale,~C.~J., Snively,~J.~B., Hickey,~M.~P., Ali,~C.~J. 2014
Thermospheric dissipation of upward propagating gravity wave
packets. Journal of Geophysical Research: Space Physics, Volume
119, Issue 5, pp. 3857-3872, doi: 10.1002/2013JA019387.
\bibitem[\protect\citeauthoryear{Hedlin at. al.}{2014}]{Hedlin2014}
Hedlin,~Michael~A.~H.; Drob,~Douglas~P. 2014. Statistical
characterization of atmospheric gravity waves by seismoacoustic
observations. Journal of Geophysical Research: Atmospheres, Volume
119, Issue 9, pp. 5345-5363, doi: 10.1002/2013JD021304.
\bibitem[\protect\citeauthoryear{Hickey et. al.}{1997}]{Hickey1997}
Hickey,~M.~P., Walterscheid,~R.~L., Taylor,~M.~J., Ward,~W.,
Schubert,~G.,Zhou,~Q., Garcia,~F., Kelly,~M.~C., Shepherd,~G.~G.
1997. Numericalsimulations of gravity waves imaged over Arecibo
during the 10-dayJanuary, campaign. J. Geophys. Res. 102 (A6),
11475-11490.
\bibitem[\protect\citeauthoryear{Hickey et. al.}{1998}]{Hickey1998}
Hickey,~M.~P., Taylor,~M.~J., Gardner,~C.~S., Gibbons,~C.~R. 1998.
Full-wavemodeling of small-scale gravity waves using Airborne
Lidar andObservations of the Hawaiian Airglow (ALOHA-93) O(1S)
images andcoincident Na wind/temperature lidar measurements. J.
Geophys. Res.103 (D6), 6439-6454.
\bibitem[\protect\citeauthoryear{Hines}{1960}]{Hines1960}
Hines,~C.~O. 1960. Internal atmospheric gravity waves at
ionospheric heights, Can. J. Phys., 38, 1441-1481.
\bibitem[\protect\citeauthoryear{Idrus et. al.}{2013}]{Idrus2013}
Idrus, Intan Izafina; Abdullah, Mardina; Hasbi, Alina Marie;
Husin, Asnawi; Yatim, Baharuddin. 2013. Large-scale traveling
ionospheric disturbances observed using GPS receivers over
high-latitude and equatorial regions. Journal of Atmospheric and
Solar-Terrestrial Physics, Volume 102, p. 321-328, doi:
10.1016/j.jastp.2013.06.014.
\bibitem[\protect\citeauthoryear{Kshevetskii and Gavrilov}{2005}]{Kshevetskii2005}
Kshevetskii~S.~P., Gavrilov~N.~M. 2005. Vertical propagation,
breaking and effects of nonlineargravity waves in the atmosphere.
Journal ofAtmospheric and Solar-Terrestrial Physics. V.67. P.
1014-1030.
\bibitem[\protect\citeauthoryear{Lindzen}{1970}]{Lindzen1970}
Lindzen,~R.~S. 1970. Internal gravity waves in atmospheres with
realistic dissipation and temperature part I. Mathematical
development and propagation of waves into the thermosphere,
Geophysical \& Astrophysical Fluid Dynamics, vol. 1, issue 3, pp.
303-355.
\bibitem[\protect\citeauthoryear{Lindzen}{1971}]{Lindzen1971_a}
Lindzen,~R.~S. 1971. Internal gravity waves in atmospheres with
realistic dissipation and temperature part III. Daily variations
in the thermosphere, Geophysical and Astrophysical Fluid Dynamics,
1971, vol. 2, Issue 1, pp.89-121
\bibitem[\protect\citeauthoryear{Lindzen and Blake}{1971}]{Lindzen1971_b}
Lindzen,~R.~S. and Blake,~D. 1971. Internal gravity waves in
atmospheres with realistic dissipation and temperature part II.
Thermal tides excited below the mesopause, Geophysical Fluid
Dynamics, 2:1, 31-61, DOI: 10.1080/03091927108236051.
\bibitem[\protect\citeauthoryear{Lindzen}{1981}]{Lindzen1981}
Lindzen~R.~S. Turbulence and stress owing to gravity wave and
tidal breakdown, J. Gephys. Res. V. 86. P. 9707-9714.
\bibitem[\protect\citeauthoryear{Luke}{1975}]{Luke}
Luke,~Y.~L. 1975. Mathematical functions and their approximations.
Academic Press. 584.
\bibitem[\protect\citeauthoryear{Lyons and Yanowitch}{1974}]{Lyons}
Lyons,~P., Yanowitch,~M. 1974. Vertical oscillatins in a viscous
and thermally conducting isotermal atmosphere. J. Fluid. Mech.
\bibitem[\protect\citeauthoryear{Ostashev}{1997}]{Ost}
Ostashev,~V.~E. 1997. Acoustics in Moving Inhomogeneous Media. E
\& FN Spon, London 259p.
\bibitem[\protect\citeauthoryear{Pierce and Posey}{1970}]{Pierce1970}
Pierce~A.~D., Posey~J.~W. 1970. Theoretical predictions of
acoustic-gravity pressure waveforms generated by large explosions
in the atmosphere, Air Force Camb. Res. Lab., AFCRL-70-0134.
\bibitem[\protect\citeauthoryear{Pierce at. al.}{1971}]{Pierce1971}
Pierce~A.~D., Posey~J.~W., Illiff~E.~F. 1971. Variation of nuclear
explosion generated acoustic-gravity wave forms with burst height
and with energy yield, J. Geophys. Res., 76(21), 5025-5041.
\bibitem[\protect\citeauthoryear{Ponomarev et. al.}{2006}]{Pon}
Ponomarev,~E.~A., Rudenko,~G.~V., Sorokin,~A.~G.,
 Dmitrienko,~I.~S.,  Lobycheva,~I.~Yu., Baryshnikov,~A.~K. 2006. Using the normal-mode method of probing the infrasonic
propagation for purposes of the comprehensive nuclear-test-ban
treaty. Journal of Atmospheric and Solar-Terrestrial Physics
\textbf{68}, 599-614.
\bibitem[\protect\citeauthoryear{Rudenko}{1994a}]{Rudenko_a}
Rudenko,~G.~V. 1994a. Linear oscillatins in a viscous and
heat-conducting isothermal atmosphere: Part 1. Atmospheric and
Oceanic Physics. \textbf{30}, No 2, 134-143.
\bibitem[\protect\citeauthoryear{Rudenko}{1994b}]{Rudenko_b}
Rudenko,~G.~V. 1994b. Linear oscillatins in a viscous and
heat-conducting isothermal atmosphere: Part 2. Atmospheric and
Oceanic Physics. \textbf{30}, No 2, 144-152.
\bibitem[\protect\citeauthoryear{Shibata and Okuzawa}{1983}]{Shibata1983}
Shibata,~T., Okuzawa,~T. 1983. Horizontal velocity dispersion of
medium-scale travelling ionospheric disturbances in the F-region.
Journal of Atmospheric and Terrestrial Physics (ISSN 0021-9169),
vol. 45, Feb.-Mar., p. 149-159.
\bibitem[\protect\citeauthoryear{Snively and Pasko}{2003}]{Snively2003}
Snively,~J.~B., Pasko,~V.~P. 2003. Breaking of
thunderstorm-generated gravity waves as a source of short-period
ducted waves at mesopause altitudes. Geophys. Res. Lett. 30 (24),
2254, doi:10.1029/2003GL018436.
\bibitem[\protect\citeauthoryear{Snively and Pasko}{2005}]{Snively2005}
Snively,~J.~B., Pasko,~V.~P. 2005. Antiphase OH and OI airglow
emissions induced by a short-period ducted gravity wave. Geophys.
Res. Lett. 32, L08808, doi:10.1029/2004GL022221.
\bibitem[\protect\citeauthoryear{Snively et. al}{2007}]{Snively2007}
Snively,~J.~B., Pasko,~V.~P., Taylor,~M.~J., Hocking,~W.~K. 2007.
Doppler ductingof short-period gravity waves by midlatitude tidal
wind structure. J. Geophys. Res. 112, A03304,
doi:10.1029/2006JA011895.
\bibitem[\protect\citeauthoryear{Vadas}{2005}]{Vadas}
Vadas,~S.~L., Fritts,~D.~C. 2005. Thermospheric responses to
gravity waves: Influences of increasing viscosity and thermal
diffusivity. J. Geophys. Res., 110, D15103,
doi:10.1029/2004JD005574.
\bibitem[\protect\citeauthoryear{Vadas and Liu}{2009}]{Vadas2009}
Vadas, Sharon ~L., Liu,~Han-li. 2009. Generation of large-scale
gravity waves and neutral winds in the thermosphere from the
dissipation of convectively generated gravity waves. Journal of
Geophysical Research, Volume 114, Issue A10, CiteID A10310, doi:
10.1029/2009JA014108.
\bibitem[\protect\citeauthoryear{Vadas and Nicolls}{2012}]{Vadas2012}
Vadas,~S.~L., Nicolls,~M.~J. 2012. The phases and amplitudes of
gravity waves propagating and dissipating in the thermosphere:
Theory. Journal of Geophysical Research, Volume 117, Issue A5,
CiteID A05322, doi: 10.1029/2011JA017426.
\bibitem[\protect\citeauthoryear{Walterscheid and Schubert}{1990}]{Walterscheid1990}
Walterscheid, R.L., Schubert, G. 1990. Nonlinear evolution of an
upwardpropagating gravity wave: overturning, convection,
transience andturbulence. J. Atmos. Sci. 47 (1), 101-125.
\bibitem[\protect\citeauthoryear{Walterscheid at. al.}{2001}]{Walterscheid2001}
Walterscheid,~R.~L., Schubert,~G., Brinkman,~D.~G. 2001.
Small-scale gravitywaves in the upper mesosphere and lower
thermosphere generated by deep tropical convection. J. Geophys.
Res. 106 (D23), 31825-31832.
\bibitem[\protect\citeauthoryear{Yu and Hickey}{2007a}]{Yu2007_a}
Yu,~Y., Hickey,~M.~P. 2007a. Time-resolved ducting of atmospheric
acousticgravitywaves by analysis of the vertical energy flux.
Geophys. Res.Lett. 34, L02821, doi:10.1029/2006GL028299.
\bibitem[\protect\citeauthoryear{Yu and Hickey}{2007b}]{Yu2007_b}
Yu,~Y., Hickey,~M.~P. 2007b. Numerical modeling of a gravity wave
packet ductedby the thermal structure of the atmosphere. J.
Geophys. Res. 112, A06308, doi:10.1029/2006JA012092,.
\bibitem[\protect\citeauthoryear{Yu and Hickey}{2007c}]{Yu2007_c}
Yu,~Y., Hickey,~M.~P. 2007c. Simulated ducting of high-frequency
atmosphericgravity waves in the presence of background winds.
Geophys. Res. Lett. 34, L11103, doi:10.1029/2007GL029591.
\bibitem[\protect\citeauthoryear{Yu at. al.}{2009}]{Yu2009}
Yu,~Y., Hickey,~M.~P., Liu,~Y. 2009. A numerical model
characterizing internal gravity wave propagation into the upper
atmosphere.Advances in Space Research, Volume 44, Issue 7, p.
836-846.,doi: 10.1016/j.asr.2009.05.014.
\bibitem[\protect\citeauthoryear{Yanowitch}{1967a}]{Yanowitch_a}
Yanowitch,~M. 1967a. Effect of viscosity on oscillatins of an
isotermal atmosphere. Can. J. Phys. \textbf{45}, 2003-2008.
\bibitem[\protect\citeauthoryear{Yanowitch}{1967b}]{Yanowitch_b}
Yanowitch,~M. 1967b. Effect of viscosity on gravity waves
 and apper boundary conditions. J. Fluid. Mech. \textbf{29}, Part
 2, 209-231.
 \end{thebibliography}
\end{document}